\documentclass[11pt,a4paper]{article}
\pdfoutput=1 
\usepackage[colorlinks=true, linkcolor=black!50!blue, urlcolor=blue, citecolor=blue, anchorcolor=blue]{hyperref}
\usepackage[font=small,labelfont=bf,margin=0mm,labelsep=period,tableposition=top]{caption}
\usepackage[a4paper,top=3cm,bottom=2.5cm,left=2.5cm,right=2.5cm,bindingoffset=0mm]{geometry}
\usepackage[T1]{fontenc} 

\usepackage[utf8]{inputenc}
\usepackage[english]{babel}

\usepackage{pifont}
\pdfoutput=1
\usepackage{graphicx,mathtools}
\usepackage{epsfig,cite}
\usepackage{amssymb}
\usepackage{amsmath}
\usepackage{color}
\usepackage{amstext,alltt,setspace}
\usepackage{amsbsy}
\usepackage{array}
\usepackage{booktabs, makecell}
\usepackage{hyperref}
\usepackage{slashed}
\usepackage{url}
\usepackage{geometry}
\usepackage{amssymb,amsthm,cancel,hyperref,graphicx,xcolor}
\usepackage{picinpar,graphicx,xy}
\usepackage[subnum]{cases}
\usepackage{booktabs}
\usepackage{mathrsfs}
\usepackage{amsfonts}
\usepackage{latexsym}
\usepackage{booktabs,graphicx,hyperref,epsfig,multirow}
\usepackage{bm}
\usepackage{comment}
\newcommand{\di}{\mathrm{d}}
\newcommand{\M}{\Bar{M}}
\newcommand{\lnM}{\ln{\M}}
\newcommand{\N}{\Bar{N}}
\newcommand{\lnN}{\ln{\N}}
\newcommand{\C}{\Tilde{C}}

\newcommand{\code}[1]{\texttt{#1}}

\def\bea{\begin{eqnarray}}
\def\eea{\end{eqnarray}}
\def\beq{\begin{equation}}
\def\eeq{\end{equation}}
\def\ba{\begin{eqnarray}}
\def\ea{\end{eqnarray}}
\def\be{\begin{equation}}
\def\ee{\end{equation}}
\definecolor{darkgreen}{HTML}{008000}
\newcommand{\sss}{\scriptscriptstyle\rm}
\newcommand{\muf}{\mu_{\rm\sss F}}
\newcommand{\mur}{\mu_{\rm\sss R}}

\newcommand{\as}{\alpha_s}

\def\({\left(}
\def\){\right)}
\def\[{\left[}
\def\]{\right]}
\def    \hepph  #1 {{\tt hep-ph/#1}}
\def    \hepex  #1 {{\tt hep-ex/#1}}
\long\def\symbolfootnote[#1]#2{\begingroup%
\def\thefootnote{\fnsymbol{footnote}}\footnote[#1]{#2}\endgroup}

\def\lapprox{\lower .7ex\hbox{$\;\stackrel{\textstyle <}{\sim}\;$}}
\def\gapprox{\lower .7ex\hbox{$\;\stackrel{\textstyle >}{\sim}\;$}}

\renewcommand{\(}{\left(}
\renewcommand{\)}{\right)}

\newcommand{\Ca}{C_{\rm\sss A}}
\newcommand{\Cf}{C_{\rm\sss F}}
\newcommand{\Nf}{n_f}

\graphicspath{{Images/}}
\usepackage{physics}

\begin{document}
\begin{flushleft}
\begin{figure}[h]
\includegraphics[width=.2\textwidth]{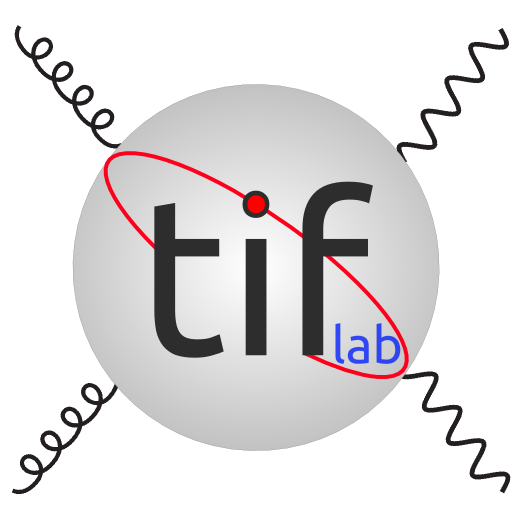}
\end{figure}
\end{flushleft}
\vspace{-5.0cm}
\begin{flushright}
TIF-UNIMI-2025-25 \\
DESY-26-004
\end{flushright}

\vspace{2.0cm}

\begin{center}
{\Large \bf Threshold resummation of Semi-Inclusive Deep-Inelastic Scattering}
\end{center}

\vspace{1.3cm}

\begin{center}
Stefano Forte$^{1,2}$, Giovanni Ridolfi$^3$, Francesco Ventola$^{1,4}$ \\
\vspace{.3cm}
{\it
{}$^1$Tif Lab, Dipartimento di Fisica, Universit\`a di Milano and \\
{}$^2$INFN, Sezione di Milano, Via Celoria 16, I-20133 Milano, Italy\\
{}$^3$Dipartimento di Fisica, Universit\`a di Genova and \\
INFN, Sezione di Genova, Via Dodecaneso 33, I-16146 Genova, Italy \\
{}$^4$ II. Institut für Theoretische Physik, Universit\"at Hamburg,\\
Luruper Chaussee 149, D-22761 Hamburg, Germany\footnote{Present address}
}
\vspace{0.9cm}


{\bf \large Abstract}
\end{center}

We derive threshold resummation of semi-inclusive deep-inelastic scattering
(SIDIS), by building upon previous results by
some of us for the resummation of the Drell-Yan process at fixed
rapidity, which is related to SIDIS by crossing. We consider both a
double-soft limit, in which both the Bjorken and the fragmentation
scaling variables tend to their threshold value, and single soft limits
in which either of them does. We show that in the former limit only
soft radiation contributes, and in the latter limit only collinear
radiation, and we derive resummed expressions for the coefficient
functions in all cases. We determine explictly resummation
coefficients in the nonsinglet channel up to next-to-next-to-leading
log by comparing to recent fixed next-to-next-to-leading order
results. Expanding out the single-soft resummation we reproduce recent
next-to-leading 
power  results.

\clearpage

\tableofcontents 

\section{Introduction}

Interest in Semi-Inclusive Deep Inelastic Scattering (SIDIS) has been
growing recently, in view of physics at the forthcoming Electron-Ion
Collider (EIC)~\cite{AbdulKhalek:2022hcn}, and indeed two independent groups
have recently computed coefficient functions for this process up to
next-to-next-to-leading
order (NNLO)~\cite{Goyal:2023zdi,Bonino:2024qbh,Goyal:2024emo,Bonino:2025qta}. It was pointed out
long ago~\cite{Sterman:2006hu}  that, because SIDIS is related by
crossing to Drell-Yan 
production, threshold resummation can be achieved for the former by
using well-known results for the latter process.

Of course, SIDIS coefficient functions depend on a pair of scaling
variables: the Bjorken 
variable $x$, physically related to the fraction of the incoming hadron
momentum carried by the incoming parton, and the fragmentation
variable $z$, related to the fraction of the final state parton that
fragments into the final-state hadron. The crossing relation therefore
relates it to the rapidity distribution of the Drell-Yan process, that
can also be parametrized by two scaling variables, $x_1$ and $x_2$,
related to the momentum fractions of the incoming partons.

The threshold resummation obtained by crossing in
Ref.~\cite{Sterman:2006hu} corresponds to the threshold resummation that
was derived long ago in Ref.~\cite{Catani:1989ne} for the Drell-Yan
rapidity distribution in the limit in which both scaling variables
tend to their threshold value. In the case of Drell-Yan this 
is the limit in which the available center-of-mass
energy tends to the threshold value needed to produce the desired
final state, namely the final-state gauge boson mass.  This
resummation for Drell-Yan has been extended recently up
next-to-next-to-leading log (NNLL) 
matched to NNLO in Ref.~\cite{Banerjee:2018vvb}, and next-to-leading
power  in Ref.~\cite{Ravindran:2022aqr}. Exploiting crossing, these
results have been used to derive the corresponding resummation for
SIDIS at increasingly high logarithmic accuracy and then also to
next-to-leading power in
Refs.~\cite{Anderle:2012rq,Abele:2021nyo,Abele:2022wuy,Goyal:2025bzf}.

The resummation of the Drell-Yan rapidity distribution has been
extended to the case in which only one of the scaling variables tends
to its threshold value using SCET techniques in
Refs.~\cite{Lustermans:2019cau,Mistlberger:2025lee}, and more recently by
us~\cite{de_ros} using the renormalization-group based approach to
soft resummation in direct QCD of Ref.~\cite{Forte:2002ni}. Here we
  show that using techniques very similar to those of
  Ref.~\cite{de_ros} it is possible to similarly perform resummation
  of SIDIS in the limit in which the Bjorken variable tends to its
  soft limit, but for fixed values of the fragmentation variable, or
  conversely.

  In the case of Drell-Yan these single soft limits correspond to the
  case in which  the final-state gauge boson is boosted by a fixed
  amount
  along the
  direction of one of the two incoming partons in the partonic center
  of mass frame, and the energy tends to the minimum value which is
  necessary in order to produce this configuration. This corresponds
  to the situation in which only the collinear radiation which is
  necessary in order to produce the required boost survives. In the
  case of SIDIS, as we shall show, the two soft limits similarly correspond to the case
  in which only radiation collinear to either the incoming or the
  fragmenting parton is allowed, and this is why the same resummation
  formalism applies.

  We  first derive a general resummation formalism by adapting
  to this case the formalism of Ref.~\cite{de_ros}. This essentially
  amounts to deriving the  kinematic configurations that
  correspond to the various soft limit and, as in our previous
  work~\cite{Forte:2002ni,Forte:2021wxe,de_ros}, determine the relevant
  soft scales through a phase-space analysis.  The resummation of the
  dependence on the soft scale then follows using the argument of
  Ref.~\cite{de_ros}, and the resummation coefficients can be
  determined by comparison to the fixed order. We will determine them
  explicitly  in the
  nonsinglet channel up to NNLL accuracy by matching to the fixed
  order results of Ref.~\cite{Bonino:2024qbh,Bonino:2025qta}, thereby
  also obtaining a nontrivial consistency check of the resummation. 

\section{Kinematics and the soft limit}
\label{SIDIS_kinematics}

We consider resummation of SIDIS coefficient functions. Because we
will match our resummed results to the fixed order results of
Refs.~\cite{Bonino:2024qbh,Bonino:2025qta} we follow notation and
conventions of these references. The measured cross-section for the
process
\begin{equation}\label{eq:2to2}
    H(P_1) + G(q) \xrightarrow{} h(P_2)  + X \ ,
\end{equation}
where $G$ is an off-shell gauge boson, is
parametrized by structure functions $\mathcal{F}_F^h(\xi,\zeta,Q^2)$,
with $F=L,\,T,\,3$ and
\begin{align}
    &Q^2 = -q^2 , \\
    &\xi = \frac{Q^2}{2P_1 \cdot q}  \\
    &\zeta = \frac{P_1 \cdot P_2}{P_1 \cdot q}. 
 \end{align}
The
structure functions can be factorized in terms of parton distributions
$f_i$ and fragmentation functions $D_j^h$ as
\begin{equation}\label{SIDIS.SF}
    \mathcal{F}^h_F(\xi,\zeta,Q^2)= \sum_{ij}\int_\xi^1\frac{dx}{x}\int_\zeta^1 \frac{dz}{z} f_i\left(\frac{\xi}{x},\mu_1^2\right) 
    C_{ij}^F\left(x,z,\as(\mur^2), \frac{\mur^2}{Q^2},\frac{\mu_1^2}{Q^2},\frac{\mu_2^2}{Q^2}\right) D_j^h\left(\frac{\zeta}{z},\mu_2^2\right).
\end{equation}

The kinematics of the partonic subprocess
\begin{equation}
    \label{eq:part_process}
    \mathcal{Q}_1(p_1) + G(q) \xrightarrow{} \mathcal{Q}_2(p_2) + X(k) \ ,
\end{equation}
is parametrized by the scaling variables
\begin{align}\label{eq:xdef}
    &{x} = \frac{Q^2}{2p_1 \cdot q} , \\ \label{eq:zdef} 
    &{z} = \frac{p_1 \cdot p_2}{p_1 \cdot q}.
\end{align}
We will henceforth focus on the partonic cross-section, and 
study the partonic process Eq.~(\ref{eq:part_process}), which can be
thought of as a $2\to2$ process, parametrized by the  masses
of the incoming gauge boson, $q^2=-Q^2$; the invariant mass
of the system $X$ that recoils
against the final-state parton, $k^2$; and the two Mandelstam
invariants
\begin{align}
\label{eq:sdef}
 s&=(p_1+q)^2=Q^2\frac{1-x}{x}\\
\label{eq:tdef}
t&=(p_1-p_2)^2=-Q^2\frac{z}{x}.
\end{align}

Resummation is performed in Mellin space, where the double convolution
of Eq.~(\ref{SIDIS.SF}) reduces to an ordinary product:
\begin{equation}\label{SIDIS.SF.ELLIN}
 {\mathcal{F}}^h_F(N,M,Q^2) = \sum_{ij} {f}_i(N,\mu^2_1) 
    C^F_{ij}\left(N,M,\as(\mur^2),\frac{\mur^2}{Q^2},\frac{\mu^2_1}{Q^2},\frac{\mu^2_2}{Q^2}\right) D_j^{h}(M,\mu^2_2)\,,
\end{equation}
where
\begin{equation}
 {\mathcal{F}}^h_F (N,M,Q^2) = \int_0^1 d\xi\, \xi^{N-1} \int_0^1 d\zeta\, \zeta^{M-1}  \mathcal{F}^h_F\left(\xi,\zeta,Q^2\right),
\end{equation}
and
\begin{align}
    &f_i(N,\muf^2) = \int_0^1 dx\, x^{N-1}f_i(x,\muf^2), 
    \\
    &D_j^{h}(M,\muf^2) = \int_0^1 dz\, z^{M-1}D_j^{h}(z,\muf^2), 
    \\
\label{SIDIS.CF.MELLIN}
    &C^F_{ij}\left(N,M,\as(\mur^2),\frac{\mur^2}{Q^2},\frac{\mu^2_1}{Q^2},\frac{\mu^2_2}{Q^2}\right)= \int_0^1 dx\, x^{N-1} \int_0^1 dz\,z^{M-1} \ 
    C_{ij}^F\left(x,z,\as(\mur^2), \frac{\mur^2}{Q^2},\frac{\mu^2_1}{Q^2},\frac{\mu^2_2}{Q^2}\right).
\end{align}
By abuse of notation we denote the Mellin transform of cross sections, parton distribution functions and fragmentation functions with the
same symbol as the untransformed quantities. This should not cause any confusion, as long as the arguments of the various functions are explicitly indicated.

\subsection{Soft limits}
\label{sec:softlimitsgr}
Inclusive DIS is a $2\to1$ process with one massive object,
and thus it is parametrized by two
invariants, that can be chosen as $Q^2$ and the center-of-mass energy
$s$ Eq.~(\ref{eq:sdef}), or equivalently the scaling variable $x$
Eq.~(\ref{eq:xdef}).  The soft limit is defined as that in which
$s\to0$, so $x\to1$. Because SIDIS is the $2\to2$ process of
Eq.~(\ref{eq:2to2}), with two
massive objects $G$ and $X$,  we have two
more invariants, that we may choose as $k^2$, the invariant mass of the
system that recoils against the final-state parton, and $t$
Eq.~(\ref{eq:tdef}). We may then define several distinct soft limits,
corresponding to different physical limits of both the partonic and
the hadronic process. In order to understand them, we first study the
kinematic bounds on the invariants of the  process.

The kinematic bounds for the Mandelstam invariant $t$ can be easily worked out in the center-of-mass frame, where
\begin{align}
&p_1=(E,0,0,E)
\\
&q=(\sqrt{E^2-Q^2},0,0,-E)
\\
&p_2=(E',0,E'\sin\theta,E'\cos\theta)
\\
&k=(\sqrt{{E'}^2+k^2},0,-E'\sin\theta,-E'\cos\theta)
\label{eq:radmom}
\end{align}
with $k^2\ge 0$. In this frame
\begin{equation}
\sqrt{s}=E+\sqrt{E^2-Q^2}=E'+\sqrt{{E'}^2+k^2},
\end{equation}
which give $E$ and $E'$ as functions of invariants:
\begin{equation}
E=\frac{s+Q^2}{2\sqrt{s}};\qquad E'=\frac{s-k^2}{2\sqrt{s}}.
\label{eq:CMenergies}
\end{equation}
As a consequence
\begin{equation}
t=(p_1-p_2)^2=-2EE'(1-\cos\theta)=-\frac{(s+Q^2)(s-k^2)}{2s}(1-\cos\theta)
\end{equation}
and therefore
\begin{equation}
-\frac{(s+Q^2)(s-k^2)}{s}\leq t\leq 0
\label{eq:tbounds}
\end{equation} 
when $0\leq \theta\leq\pi$.  The minimum value of $t$ is attained when
$k^2=0$, hence for fixed $s$
\begin{align}\label{eq:tbound}
&  t^{\rm min} \le t\le 0,\\ \label{eq:tmin}
  &\quad  t^{\rm min}= - \left(s+Q^2\right)=-\frac{Q^2}{x}.
  \end{align}
The kinematic limit for $k^2$ can be determined
substituting Eqs.~(\ref{eq:sdef},\ref{eq:tdef}) in Eq.~(\ref{eq:tbounds}). We find
\begin{equation}
Q^2\frac{z}{x}\leq \frac{Q^2}{x}-\frac{k^2}{1-x}.
\end{equation}
whence
\begin{equation}
k^2\leq Q^2\frac{(1-x)(1-z)}{x}.
\end{equation}

The soft limit is in general that in which all radiation is suppressed, so
the process tends to the leading-order process. At leading order
$p_1+q$ is the momentum of an on-shell parton, hence
Eq.~(\ref{eq:sdef}) implies that $x=1$, while  $p_1-p_2=-q$, hence
Eq.~(\ref{eq:tdef}) implies that $z=x$. We can consequently define: a
double soft limit, as the limit in which both $x\to1$ and $z\to1$; and
two single soft limits, in which $x\to1$ at fixed $z$ or $z\to1$ at
fixed $x$. Clearly, in terms of Mandelstam invariants the limit
$x\to1$ at fixed $z$ corresponds to the case in which $s$ tends to its
minimum, $s\to0$ but  $t$ takes some fixed value in the limit, while
the limit
$z\to1$ at fixed $x$ corresponds to the case in which $t$ tends to its
minimum Eq.~(\ref{eq:tmin}), $t\to t^{\rm min}$ but  $s$ is fixed.

The
physical meaning of these limits can be understood in terms of the
allowed radiation in each case.
In the limit $x\to 1$ from Eq.~(\ref{eq:xdef})
we have $(p_1+q)^2\to 0$. Using momentum conservation,
\begin{equation}\label{eq:x1cond}
(p_1+q)^2=(p_2+k)^2=k^2+2p_2\cdot k\to 0,
\end{equation}
which is achieved  when $k$ is either soft, $k\to 0$, or massless and collinear to
$p_2$,  $k\to a p_2$.
In the limit $z\to 1$ instead from Eq.~(\ref{eq:zdef}) we have
$p_1\cdot p_2\to p_1\cdot q$, so using again momentum conservation
\begin{equation}\label{eq:z1cond}
  p_1\cdot(p_1-k)=-p_1\cdot k\to 0,
\end{equation}
which is  achieved  when $k$ is either soft, $k\to 0$, or massless and collinear to
$p_1$,  $k\to b p_1$.

It follows that the double soft limit corresponds to the case in which
$k$ is soft, while each of the two single soft limits corresponds to
$k$ being either collinear to $p_1$ or to $p_2$.
In particular the two limits for the non-soft scaling variable are
respectively given by
\begin{align}
\lim_{k\to a p_2} z&=\lim_{k\to ap_2} \frac{p_1\cdot p_2}{p_1\cdot
  p_2+p_1\cdot k }=\frac{1}{1+a},\\
\lim_{k\to b p_1}x&=\lim_{k\to bp_1}\frac{-(p_2+k-p_1)^2}{2p_1\cdot(p_2+k)}=1-b.
\end{align}

As pointed out in Ref.~\cite{Sterman:2006hu} and as mentioned in the
introduction, SIDIS is the crossed 
process to Drell-Yan, and indeed these soft limits are in one-to-one
correspondence with the soft limits of the (partonic) Drell-Yan
rapidity distribution, discussed in Ref.~\cite{de_ros}. Specifically,
one can view the Drell-Yan partonic process at nonzero rapidity as  the
$2\to 2$ process
\begin{equation}
    \label{eq:DYdef}
    \mathcal{Q}_1(\bar p_1) + \mathcal{Q}_2(\bar p_2)  \xrightarrow{}  G(\bar q) + X(\bar k),
\end{equation}
in which two incoming massless partons $\mathcal{Q}_i$ produce a gauge boson
with mass $q^2=Q^2$ and a partonic system with invariant
mass $k^2$ that recoils against the gauge boson in order for it to have
nonzero rapidity.

The crossing relations between the two sets of momenta are
\begin{align}
    & p_1 \leftrightarrow \bar p_1  \\
    & p_2 \leftrightarrow -\bar p_2  \\
    & k \leftrightarrow \bar k   \\
    & q \leftrightarrow - \bar q 
\end{align}
and the relation between Mandelstam invariants is
\begin{equation}
    s \leftrightarrow t .
\end{equation}

The scaling variables for the Drell-Yan process can be defined by
choosing the partonic center-of-mass frame, in which
\begin{align}
    &\bar p_1=\frac{\sqrt{s}}{2}(1,0,0,1) \\
  &\bar p_2=\frac{\sqrt{s}}{2}(1,0,0,-1),
  \end{align}
and then letting
\begin{align}\label{EQ.inv.vars.DY}
    &x_1 x_2 = \frac{Q^2}{s}\\
    & \frac{x_1}{x_2} =e^{2y},
\end{align}
where $y$ is the rapidity of the gauge boson.

The double soft limit $x_1\to 1$, $x_2\to1$ then corresponds to the
case in which $s\to Q^2$, so all radiation is suppressed and $k$ is
soft.
The two single soft limits $x_1\to 1$ with $x_2$ fixed and conversely
correspond to the case in which the rapidity of the gauge boson is
fixed in the limit, and $s$ reaches the minimum value compatible with
this requirement. This means that the transverse momentum of the
gauge boson then tends to zero, and consequently in the center-of-mass
frame $k$ is massless and collinear to the incoming partons,
specifically collinear to $p_1$ or  to $p_2$ 
according to whether $x_1$ is fixed and $x_2\to1$ or conversely. 
The physical interpretation of the two limits is thus the same for
SIDIS and Drell-Yan.

\subsection{The phase space measure in soft limits}
\label{phase space}
In this Section we work out an expression for the phase space measure
for the SIDIS process with the emission of $n$ massless
partons. Following the approach of Refs.~\cite{Forte:2002ni,Forte:2021wxe,de_ros}, the goal is
to show that in the soft limit all the dependence on the scaling
variable goes through a dimensionful combination that corresponds to
the soft scale of the process.
The phase space measure can be written as
\begin{equation}
    \label{eq:psn+1}
    d\phi_{n+1}(p_1+q;\,p_2,\,k_1, \dots, k_2) = \int_0^{k^2_{\rm max}} \frac{dk^2}{2\pi}\,d\phi_2(p_1+q;\,p_2,\,k)\,d\phi_n(k;\,k_1, \dots,k_n)
\end{equation}
where 
\begin{equation}
k^2_{\rm max}=Q^2\frac{(1-x)(1-z)}{x},
\label{eq:k2max}
\end{equation}
as shown in Sect.~\ref{sec:softlimitsgr}.We first compute the two-body phase space in $d=4-2\epsilon$ dimensions
\begin{align}
    d\phi_2(p_1+q;p_2,k) &=
    \frac{d^{d-1}k}{(2\pi)^{d-1}2k^0}\frac{d^{d-1}p_2}{(2\pi)^{d-1}2p_2^0}(2\pi)^d \ \delta^{(d)}\left(p_1+q-p_2-k\right) 
\nonumber \\
    &= \frac{1}{4(2\pi)^{d-2}}\frac{d^{d-1}p_2}{p_2^0k^0} \ \delta\left(p_1^0+q^0-p_2^0-k^0\right).
\end{align}
We adopt the center-of-mass frame defined in the previous subsection, and we switch to $(d-1)$-dimensional polar coordinates for $\vec p_2$. After performing an irrelevant azimuthal integration, we get
\begin{equation}
d\phi_2(p_1+q;p_2,k) =\frac{1}{4(2\pi)^{d-2}} \frac{2\pi^{\frac{d-2}{2}}}{\Gamma\left(\frac{d-2}{2}\right)}\sin^{d-4}\theta
d\cos\theta
\frac{|\vec p_2|^{d-3}d|\vec p_2|}{\sqrt{|\vec p_2|^2+k^2}} \delta\left(\sqrt{s}-|\vec p_2|-\sqrt{|\vec p_2|^2+k^2}\right).
\end{equation}
The energy delta function
\begin{equation}
\delta\left(\sqrt{s}-|\vec p_2|-\sqrt{|\vec p_2|^2+k^2}\right)=\frac{\sqrt{|\vec p_2|^2+k^2}}{\sqrt{s}}\delta\left(|\vec p_2|-\frac{s-k^2}{2\sqrt{s}}\right)
\end{equation}
can be used to perform the $|\vec p_2|$ integration. The only remaining independent variable, the scattering angle $\theta$, can be traded for $z$:
\begin{align}
&z=\frac{p_1\cdot p_2}{p_1\cdot q}=\frac{s-k^2}{2s}(1-\cos\theta)
\\
&\sin\theta=\frac{2\sqrt{s}}{s-k^2}\sqrt{z[s(1-z)-k^2]}.
\end{align}
Hence, setting as usual $d=4-2\epsilon$,
\begin{equation}
d\phi_2(p_1+q;p_2,k) =\frac{1}{8\pi} \frac{(4\pi)^\epsilon}{\Gamma\left(1-\epsilon\right)}
\left(\frac{s-k^2}{2\sqrt{s}}\sin\theta\right)^{-2\epsilon}dz
\end{equation}
and the phase space in Eq.~(\ref{eq:psn+1}) becomes
\begin{equation}
    \label{eq:psn+1new}
    d\phi_{n+1}(p_1+q;p_2,k_1, \dots ,k_n) = dz\,\frac{(4\pi)^\epsilon}{8\pi\Gamma(1-\epsilon)}\int_0^{k^2_{\rm max}}
    \frac{dk^2}{2\pi}\,\left(\frac{s-k^2}{2\sqrt{s}}\sin\theta\right)^{-2\epsilon}
    d\phi_n(k;k_1, \dots, k_n).
\end{equation}

We can now collect all the dimensionful dependence by setting
\begin{equation}
    k^2 = vk^2_{\rm max}; \qquad 0 \leq v \leq 1,
\end{equation}
so that
\begin{equation}
    dk^2=dv \ Q^2\frac{(1-x)(1-z)}{x}.
\end{equation}
The factor
\begin{equation}
\left(\frac{s-k^2}{2\sqrt{s}}\sin\theta\right)^2=z[s(1-z)-k^2]=k^2_{\rm max}z(1-v)
\end{equation}
is simply the squared transverse momentum of the radiated system, as one can read off Eqs.~(\ref{eq:radmom}) and~(\ref{eq:CMenergies}). We see from Eq.~(\ref{eq:k2max}) that it vanishes both in the double soft and in the single soft limits.
The phase space $d\phi_n(k; k_1, \dots , k_n)$ has the same structure
as in deep-inelastic scattering, with incoming momentum $k$.
As in  Ref.~\cite{Forte:2002ni}, it is  written in terms of a
dimensionless integration measure, with the dimensionful dependence
contained in a power of $k^2$.
Thus, following Ref.~\cite{Forte:2002ni}, we obtain
\begin{equation}
    \label{eq:ps_Dis}
    d\phi_n(k;k_1, \dots ,k_n) = 2\pi\left[\frac{N(\epsilon)}{2\pi}\right]^{n-1}(k^2)^{n-2-(n-1)\epsilon}d\Omega^{n-1}(\epsilon),
\end{equation}
where $N(\epsilon)=\frac{1}{2(4\pi)^{2-2\epsilon}}$ and
\begin{equation}
    d\Omega^{n-1}(\epsilon) = d\Omega_1 \dots d\Omega_{n-1} \int_0^1 dz_{n-1}z_{n-1}^{(n-3)-(n-2)\epsilon}(1-z_{n-1})^{1-2\epsilon}\dots\int_0^1dz_2 z_2^{-\epsilon}(1-z_2)^{1-2\epsilon} \ .
\end{equation}
The definition of the dimensionless $z_i$ variables is irrelevant here.

We are now in a position to discuss the scale dependence of the phase space
in the soft limits. As  shown in Sect.~\ref{sec:softlimitsgr}, $k^2\xrightarrow{} 0$ as either  $x$ or  $z$ approach 1.
In then follows from Eqs.~(\ref{eq:psn+1new},\ref{eq:ps_Dis}) that
the dimensionful dependence of the full phase space is entirely
contained in powers $(\Lambda^2)^{n\epsilon}$ of a soft scale 
\begin{equation}
    \label{eq:double_soft_scale}
    \Lambda^2=k^2_{\rm max} = Q^2(1-x)(1-z)[1+\mathcal{O}(1-x)].
\end{equation}
The logarithmic dependence of the cross section then comes from the
interference with poles in 
$\epsilon$ that arise when integrating over the squared amplitude with the  remaining
dimensionless measure~\cite{Forte:2002ni}.
In the double soft limit the soft scale is thus given by
Eq.~(\ref{eq:double_soft_scale}), while in each of the two single soft
limits the variables $1-z$ or $1-x$ are not soft, and thus the soft
scale can be equivalently taken to be equal to $Q^2(1-x)$ or $Q^2(1-z)$
respectively, up to a finite rescaling factor. 

\section{Resummation up to NNLL accuracy}
\label{Sec:resummation}

The main result of Sect.~\ref{SIDIS_kinematics} is that the SIDIS
structure functions depend on a single soft scale in both the
double soft and the two single soft limits. The situation is identical
to that encountered when deriving the corresponding limits of the
Drell-Yan process, discussed in Ref.~\cite{de_ros}, and the underlying
physical reason is the same, namely, the collinear nature of these
limits. Resummation can then be performed following the same argument as in
Ref.~\cite{de_ros}. We first summarize the resummed results, then
determine the resummation coefficients by comparison to the fixed
order calculation.

\subsection{The resummed coefficient function}
\label{sec:rescf}

Resummation is performed in Mellin space, where the soft scale
Eq.~(\ref{eq:double_soft_scale}) becomes~\cite{de_ros}
\begin{equation}\label{eq:dss}
  \bar\Lambda^2=\frac{Q^2}{MN}.
\end{equation}
The double soft limit corresponds then to $M\to\infty$, $N\to\infty$,
while the two single soft limits corresponds to either  $M\to\infty$
with  finite  $N$ or conversely. In the two single soft limits the
soft scale thus becomes $\frac{Q^2}{M}$ or $\frac{Q^2}{N}$ up to a
finite rescaling by the non-soft Mellin variable. Resummation is
performed by only retaining terms in Mellin space that do not vanish
in the respective limits.  

Resummed results are found by a rerun of the argument of
Ref.~\cite{de_ros}. We note that only the
transverse structure function is enhanced in the soft limit. For this reason, to simplify notations, we omit henceforth the $F$ label that distinguishes different structure functions. In the double soft limit, 
choosing
$\mur^2=\mu_1^2=\mu_2^2=Q^2$  the resummed coefficient function is
given by
\begin{align}
\label{RES.SIDIS.DS.1Int.NNLL}
&C_{qq}^{\text{ds}}\left(N,M,\as(Q^2)\right) = C_{c,qq}^{\text{ds}}\left(\as(Q^2)\right) 
\exp\int_{\frac{Q^2}{NM}}^{Q^2}\frac{dk^2}{k^2}\left[A_q(\as(k^2))\ln{\frac{N M k^2}{Q^2}}
-D_{qq}^{\text{ds}}(\as(k^2))\right],
\end{align}
where both parton indices are taken to be $q$ because in the soft limit only
the quark-quark channel is enhanced.

In Eq.~(\ref{RES.SIDIS.DS.1Int.NNLL})  the functions
$A_q$, $D_{qq}^{\text{ds}}$  and $C_{c,qq}^{\text{ds}}$ are all power series in $\as$
\begin{align}
\label{eq:acdef}
&A_q(\as)=\sum_{n=1}^\infty\left(\frac{\as}{\pi}\right)^nA_{q}^{(n)}
\\
\label{eq:dcdef}
&D_{qq}^{\text{ds}}(\as)=\sum_{n=1}^{\infty}\left(\frac{\as}{\pi}\right)^nD_{qq}^{\text{ds}(n)}
\\
\label{eq:gcdef}
& C_{c,qq}^{\text{ds}}(\as)=1+\sum_{n=1}^\infty\left(\frac{\as}{\pi}\right)^n C_{c,qq}^{\text{ds}(n)} ; 
 \end{align}
$A_q$ is the quark cusp anomalous dimension.
 Assuming knowledge of the cusp anomalous dimension up to $\mathcal{O}(\alpha^{k+1}_s)$, the fixed N$^k$LO
 result fully determines  N$^k$LL resummation; below we will determine 
 $D_{qq}^{\text{ds}}$ and
 $C_{c,qq}^{\text{ds}}$ up to $\mathcal{O}(\as^2)$ by comparison with the fixed
 order
 calculation~\cite{Goyal:2023zdi,Bonino:2024qbh,Goyal:2024emo,Bonino:2025qta}.

In the single soft limit, up to leading power accuracy,
only soft diagonal radiation is allowed from the
soft parton, which thus for the DIS process must be a quark, while all collinear radiation
is allowed from the non-soft parton. The resummed coefficient function is
consequently labeled by a quark index and a parton index $j$, that runs over all
parton flavors, for the non-soft parton. In order to ensure
consistency of the resummed result with standard perturbative
evolution, which dictates the dependence on $\mu_F^2=Q^2$ of the
coefficient function, the parton index $j$ must run over evolution
eigenstates~\cite{DeRos:2026bcv}. 
Considering for definiteness the $x\to1$, $N\to\infty$
single soft limit, the resummed result is then
\begin{equation}
    \label{RES.SIDIS.SS.1Int.NNLL}
    C^{\text{ss}}_{qj}\left(N,M,\as(Q^2)\right) = C_{c,qj}^{\text{ss}}\left(\as(Q^2),M\right)
    \exp\int_{\frac{Q^2}{N}}^{Q^2}\frac{\di k^2}{k^2}\left[A_q(\as(k^2))\ln{\frac{N k^2}{Q^2}}-D_{qj}^{\text{ss}}(\as(k^2),M)\right].
\end{equation}
The result in the $z\to1$, $M\to\infty$ limit has the same form,
but with $N\leftrightarrow M$.
The functions $D_{qj}^{\text{ss}}$ and
$C_{c,qj}^{\text{ss}}$ now depend parametrically on the non-soft variable.
In Eq.~(\ref{RES.SIDIS.SS.1Int.NNLL}) the soft scale is
\begin{equation}\label{eq:sss}
  \bar\Lambda_{\rm ss}^2=\frac{Q^2}{N}.
\end{equation}
The result can of course be rewritten by choosing any
scale   $\bar\Lambda^{\prime\,2}_{\rm ss}= k   \bar\Lambda_{\rm ss}^2 $ that
differs by a finite factor $k$, such as specifically the double soft 
 scale $\bar\Lambda^2$ Eq.~(\ref{eq:dss}).

 \subsection{Expanded resummed results}

In order to determine the resummation coefficients, we match the
resummed expression to the
fixed order calculation up to $\mathcal{O}(\alpha_s^2)$, specifically using the result of
Refs.~\cite{Bonino:2024qbh,Bonino:2025qta}. As
mentioned, whereas in the double soft limit only the quark-quark
channel is enhanced, in the single soft limit the non-soft parton can
correspond to any parton flavor. Here we will only
consider the nonsinglet contribution,  hence we only match and resum
the $\tilde C_{qq,\,{\rm NS}}$ coefficient function. We
will drop the NS subscript and implicitly denote with
$qq$ the nonsinglet quark channel.

The fixed-order Mellin-space expressions that we used are listed in
Appendix~\ref{CFSofts}. They have been obtained by us starting from
the NNLO $(x,z)$-space expressions of
Ref.~\cite{Bonino:2024qbh,Bonino:2025qta}, and performing the Mellin transform 
of the distributional contributions 
using the \code{Mathematica} packages
\code{MT}~\cite{Hoschele:2013pvt}  \code{HarmonicSums}~\cite{Ablinger:2012ufz}.

The matching procedure requires the expansion of the resummed results
in the double soft Eq.~(\ref{RES.SIDIS.DS.1Int.NNLL}) and single soft
Eq.~(\ref{RES.SIDIS.SS.1Int.NNLL}) cases. We note that the double soft
and single soft expressions have the same form when expressed in terms of a variable
\begin{equation}\label{eq:lamdef}
 \lambda=\beta_0\alpha_s(Q^2) \mathcal{L} 
\end{equation}
where $\mathcal{L}$ is a large log, though with different coefficients
$A_q^{(n)}$, $D_{ij}^{(n)}$ and $C_{ij}^{(n)}$ Eqs.~(\ref{eq:acdef}-\ref{eq:gcdef}),
and in general a different choice of the large log $\mathcal{L} $.
It is furthermore convenient to write results in terms of
modified Mellin variables~\cite{Catani:1989ne}, defined as
\begin{equation}\label{eq:mnbar}
  \N = N e^{\gamma_E};\quad  \M = M e^{\gamma_E},
\end{equation}
where  $\gamma_E$ is the Euler–Mascheroni constant, as this leads to a
simplification of the resummation coefficients.
We then take as large logs in the double soft limit
\begin{equation}\label{eq:lamds}
\mathcal{L}= \ln(\N  \M), 
\end{equation}
while in the single soft limit we give results  in terms of the
large logs $\ln\N$ or   $\ln\M$ respectively, but also in terms of
$\ln \M\N$, which of course when either $M$ or
$N$ are fixed and finite differs from them by a finite rescaling
of the argument of the log.

Performing the integral over scale, the resummed results have  the form
\begin{equation}
    \label{resum_nnll_noexp}
    C_{qq}\left(N,M,\as(Q^2)\right) = g_0(\as)\exp\left(\frac{1}{\as}g_1(\lambda)+g_2(\lambda)+\as g_3(\lambda)+\ldots\right),
\end{equation}
where
\begin{equation}
g_0(\as)=1+g_{0}^{(1)}\frac{\as}{\pi}+g_{0}^{(2)}\left(\frac{\as}{\pi}\right)^2+\ldots
\end{equation}
and
\begin{align}    \label{resum_nnll_g1}
    &g_1(\lambda) = \frac{A_q^{(1)}}{\pi \beta_0^2} (\lambda + (1-\lambda )\ln(1 - \lambda)) \,, \\ \label{resum_nnll_g2}
    &g_2(\lambda)=\frac{A_q^{(1)} b_1}{2 \pi \beta_0^2} \left(2 \lambda +\ln^2(1-\lambda )+2\ln (1-\lambda )\right)-\frac{A_q^{(2)}}{\pi ^2 \beta_0^2}(\lambda +\log (1-\lambda )) \\ 
    &\phantom{f_2(\lambda) =}\,+\frac{D_{qq}^{(1)}\ln (1-\lambda )}{\pi  \beta_0} \,, \nonumber \\  \label{resum_nnll_g3}
    &g_3(\lambda) = \frac{A_q^{(1)} b_1^2}{\pi  \beta_0^2 (1-\lambda)} \left(\frac{\lambda ^2}{2}+\frac{1}{2}\ln ^2(1-\lambda )+\lambda \ln (1-\lambda )\right) \nonumber \\ 
    &\phantom{f_3(\lambda) =}\,+\frac{A_q^{(1)} b_2}{\pi  \beta_0^2 (1-\lambda)} \left(\log (1-\lambda )-\lambda\log (1-\lambda )-\frac{\lambda ^2}{2} +\lambda\right) \nonumber \\ 
    &\phantom{f_3(\lambda) =}\,-\frac{A_q^{(2)} b_1}{\pi ^2 \beta_0^2 (1-\lambda)} \left(\frac{\lambda ^2}{2}+\lambda +\log (1-\lambda )\right)+\frac{A_q^{(3)} \lambda ^2}{2 \pi ^3 \beta_0^2 (1-\lambda )} \nonumber \\ 
    &\phantom{f_3(\lambda) =}\,-\frac{D_{qq}^{(2)} \lambda }{\pi ^2 \beta_0 (1-\lambda)} +\frac{D_{qq}^{(1)} b_1}{\pi  \beta_0 (1-\lambda)} (\lambda + \ln (1-\lambda )),
\end{align}
where $b_i=\frac{\beta_i}{\beta_0}$ for $i\ge 1$. The coefficients $\beta_0$ and $\beta_i$ are explicitly given in  Appendix~\ref{running_coupling}.

Expanding now Eq.~(\ref{resum_nnll_noexp}) up to order $\as^2$ we get
\begin{align}\label{NNLL.DS.EXPANDED}
    C_{qq}\left(N,M,\as(Q^2)\right) &=  1+\frac{\as}{\pi}\left(\frac{{A_{q}^{(1)}}}{2} \mathcal{L}^2 - D^{(1)}_{qq}\mathcal{L}+g_{0}^{(1)}\right) \nonumber \\
    &+\left(\frac{\as}{\pi}\right)^2\Bigg[ \frac{{A_{q}^{(1)}}^2}{8} \mathcal{L}^4+ \mathcal{L}^3\left(\frac{1}{6}{A_{q}^{(1)}}\pi \beta_0 - \frac{1}{2} {A_{q}^{(1)}} D_{qq}^{(1)}\right) 
 \nonumber \\
&
+\mathcal{L}^2 \left(\frac{1}{2}{A_{q}^{(1)}} g_{0}^{(1)} +\frac{1}{2} {A_{q}^{(2)}} - \frac{1}{2}\pi \beta_0 {D_{qq}^{(1)}} +\frac{1}{2}{D_{qq}^{(1)}}^2 \right)\nonumber \\
    &+\mathcal{L}\left(- {D_{qq}^{(1)}} g_{0}^{(1)}- {D_{qq}^{(2)}}\right)+g_{0}^{(2)} \Bigg]+\mathcal{O}(\as^3).
\end{align}

Resummed results will be given in both the double soft and single soft
limit,  in the latter case with two different choices of soft
scale, by determining in each case the values of the coefficients
$A_{q}^{(i)}$, $D_{qq}^{(i)}$ and $g_{0}^{(i)}$ to be used in the
  expressions Eqs.~(\ref{resum_nnll_noexp}-\ref{resum_nnll_g3}). Note
  in particular that the function $g_0$ does not in general coincide
  with the functions $C_{c}^{ds}$ and  $C_{c}^{ss}$
  Eqs.~(\ref{eq:gcdef}-\ref{RES.SIDIS.SS.1Int.NNLL}), because
  different choices of soft scale can reshuffle constant terms between
  the exponential and the unexponentiated prefactor.

\subsection{Double soft}
As mentioned in the introduction, resummation of SIDIS in the double
soft limit was already presented in Refs.~\cite{Anderle:2012rq} (NLL),
\cite{Abele:2021nyo} (NNLL) and even N$^3$LL~\cite{Abele:2022wuy},
also including next-to-leading power corrections. In these references,
resummed results were obtained using crossing
symmetry~\cite{Sterman:2006hu}, exploiting the close connection of differential
to inclusive resummation~\cite{Catani:1989ne}, and using recent higher-order
results for the inclusive resummation functions $\Delta$ and $J$ (see
e.g.~\cite{Moch:2005ba}).  We
present here resummation up to NNLL with coefficients determined by matching to the fixed order result, which sets the stage for our new
results on resummation in the single soft limit, and also provides an
independent cross-check of the consistency of the fixed order and
resummed results.

The coefficient function in the double soft limit is at NLO
\begin{equation}\label{nlo_qq_channel}
    C_{qq}^{(1)} = C_F\left(\frac{{\cal L}^2}{2}+\frac{\pi ^2}{6}-4\right)+\mathcal{O}\left(\frac{1}{N}\right)+\mathcal{O}\left(\frac{1}{M}\right)\,,\quad 
\end{equation}
and at NNLO
\begin{align}\label{nnlo_qq_channel}
    C^{(2)}_{qq}\left(N,M\right)\, &= \,
    \frac{{\cal L}^4}{8}C_F^2  + C_F {\cal L}^3 \left(\frac{\pi}{6} \,\beta_0 \right) \nonumber \\ 
    &+\,C_F\frac{{\cal L}^2}{4}\left[ C_F \left(-8+\frac{\pi^2}{3}\right)+\left(\frac{67}{18}-\frac{\pi ^2}{6}\right) C_{A}-\frac{5}{9} N_{f}\right]\nonumber\\ 
    &+\,C_F \frac{{\cal L}}{2}\left[ \left(\frac{101}{27}-\frac{7}{2}\,\zeta_3\right)\,C_A -\frac{14}{27}\,N_f\right]\nonumber\\ 
    &+\,C_F^2\left[ \frac{511}{64}-\frac{\pi^2}{16}-\frac{\pi^4}{60}-\frac{15}{4}\,\zeta_3\right]  \nonumber\\ 
    &+\,C_F C_A \left[ -\frac{1535}{192}-\frac{5\pi^2}{16}+\frac{7\pi^4}{720}+\frac{151}{36}\,\zeta_3\right] \nonumber\\ 
    & +C_F N_f\left[ \frac{127}{96}+\frac{\pi^2}{24}+\frac{\zeta_3}{18}\right] +\mathcal{O}\left(\frac{1}{N}\right)+\mathcal{O}\left(\frac{1}{M}\right)\,.
\end{align}

Comparing to  Eq.~(\ref{nlo_qq_channel}) we
immediately get the NLL coefficients
\begin{align}\label{eq:adsn}
    &A_q^{(1)}= C_F \\
    \label{eq:ddsn}
    &D_{qq}^{\text{ds}(1)} = 0\\
    \label{eq:gdsn}
    &g_{0,qq}^{\text{ds}(1)} =C_F\left(\frac{\pi^2}{6}-4\right)=C_F\left(\zeta_2-4\right),
\end{align}
where of course the value of $A_q^{(1)}$ provides a consistency check, as it
is already fixed by LL resummation, and the full NLL resummation
requires knowledge of the NLO cusp anomalous dimension, which is given
by
\begin{equation}\label{eq:nlocusp}
    A_q^{(2)} = \frac{1}{2}C_F\left[\left(\frac{67}{18}-\frac{\pi^2}{6}\right) C_{A}-\frac{5}{9} N_{f}\right] .
\end{equation}

At order $\alpha_s^2$, comparing to   Eq.~(\ref{nnlo_qq_channel}) we
reproduce Eq.~(\ref{eq:nlocusp}), and we find
\begin{align}
    &D_{qq}^{\text{ds},(2)} = \frac{1}{2}C_F\left[ \left(-\frac{101}{27}+\frac{7}{2}\,\zeta_3\right)\,C_A +\frac{14}{27}\,N_f\right] \,,\\\label{eq:g2ds}
    &g_{0,qq}^{\text{ds},(2)} = C_F^2\left[ \frac{511}{64}-\frac{\pi^2}{16}-\frac{\pi^4}{60}-\frac{15}{4}\,\zeta_3\right]  \nonumber\\ 
    &\phantom{g_{0,(\text{ds},qq)}^{(2)} =} +\,C_F C_A \left[ -\frac{1535}{192}-\frac{5\pi^2}{16}+\frac{7\pi^4}{720}+\frac{151}{36}\,\zeta_3\right] \nonumber\\ 
    &\phantom{g_{0,(\text{ds},qq)}^{(2)} =} +C_F N_f\left[ \frac{127}{96}+\frac{\pi^2}{24}+\frac{\zeta_3}{18}\right] \,,
\end{align}
which fully determine the NNLL resummed result, together with the NNLO
cusp anomalous dimension given in
Refs.~\cite{Moch:2004pa,Moch:2005ba}).
This agrees with  the results
of Ref.\cite{Abele:2021nyo} (note that  the definition of the
coefficient $D_2$ in this reference differs by a factor 2 from our own).

\subsection{Single soft}
We now turn to the new result of this paper, namely the coefficients
that determine the single soft resummation in the nonsinglet channel.
In order to perform the matching efficiently, it is useful to expand
the fixed-order results in powers of either of the large  logs
corresponding to the two single soft limits, namely
\begin{align}
    \label{CTQ2Q_singleN}
   C_{qq}(N,M) &= 1 + \frac{\as}{\pi}\left(f_2^{(1)}(M)\ln^2{\N}+ f_1^{(1)}(M)\lnN+f_0^{(1)}(M)\right)
    \nonumber \\
   &+\left(\frac{\as}{\pi}\right)^2\left(f_4^{(2)}(M)\ln^4{\N}+f_3^{(2)}(M)\ln^3{\N} \right. 
   \nonumber \\
   &\left.+f_2^{(2)}(M)\ln^2{\N}+f_1^{(2)}(M)\lnN +f_0^{(2)}(M)\right) 
   \nonumber \\
   & +\mathcal{O}\left(\frac{1}{N}\right)+\mathcal{O}(\as^3)
   \\
  \label{CTQ2Q_singleM}
   C_{qq}(N,M)&= 1 + \frac{\as}{\pi}\left(h_2^{(1)}(N)\ln^2{\M}+h_1^{(1)}(N)\lnM +h_0^{(1)}(N)\right) 
   \nonumber \\
   &+\left(\frac{\as}{\pi}\right)^2\left(h_4^{(2)}(N)\ln^4{\M}+h_3^{(2)}(N) \ln^3{\M}\right. 
   \nonumber \\
   &\left.+h_2^{(2)}(N)\ln^2{\M}+h_1^{(2)}(N)\lnM +h_0^{(2)}(N)\right) 
   \nonumber \\
   & +\mathcal{O}\left(\frac{1}{M}\right)+\mathcal{O}(\as^3).
   \end{align}
The explicit expressions of the functions $f_i^{(j)}(M)$ and
$h_i^{(j)}(N)$ are given in Appendix~\ref{CFSofts}.

\subsubsection{$x$-single soft}

Taking as expanded resummed expression Eq.~(\ref{resum_nnll_noexp}),
but now with
\begin{equation}\label{eq:lamdsx}
\mathcal{L}= \ln\N, 
\end{equation}
and comparing with the expressions of the functions  $f_i^{(j)}(M)$
given in Appendix~\ref{app:CFSoftx} we get
\begin{align}
    &A^{(1)}_{q}=C_F\,, \\ 
    &A^{(2)}_{q}=\frac{1}{2}C_F\left[\left(\frac{67}{18}-\frac{\pi ^2}{6}\right) C_{A}-\frac{5}{9} N_{f}\right]\,, \\ \label{eq:xssd1s}
    &D^{\text{xss},(1)}_{qq}(M)=\Tilde{\gamma}^{(0)}_{qq}(M)\,, \\ \label{eq:xssd2s}
    &D^{\text{xss},(2)}_{qq}(M)=D_{qq}^{\text{ds},(2)}-\pi\beta_0 F(M)+\Tilde{\gamma}^{(1),t}_{qq,\text{NS}}(M)\,, \\ 
    &g_{0,qq}^{\text{xss},(1)}(M)=g_{0,qq}^{\text{ds},(1)}+F(M) = f_0^{(1)}(M),\\ 
    &g_{0,qq}^{\text{xss},(2)}(M)=f^{(2)}_0(M).
\end{align}
Here $\Tilde \gamma^{(0)}_{qq}(M),\Tilde
\gamma^{(1),t}_{qq,\text{NS}}(M)$ are subtracted  N$^i$LO timelike nonsinglet
anomalous dimensions, obtained as  the Mellin transforms of
the N$^i$LO timelike nonsinglet splitting functions, but with contributions proportional to
$\delta(1-z)$ not included, as explained in Appendix~\ref{spNLO}.
The suffix $t$ (for timelike) is only present in the $\mathcal{O}(\as^2)$ coefficient because at leading order the timelike and spacelike splitting
functions coincide. Also,
\begin{align}
    F(M)=  \frac{1}{2} C_F\left[ S_1^2(M)+\frac{2 M^2-M-1}{M^2 (M+1)^2}+  3\left(S_2(M)-\zeta_2\right)-\frac{S_1(M)}{M (M+1)}\right]\,.
\end{align}
Substituting these coefficients in the resummed expression
Eq.~(\ref{RES.SIDIS.SS.1Int.NNLL}) gives NNLL resummation in the
$x$-single soft limit.

As mentioned, it might be convenient to rewrite this result by
choosing instead as a soft scale the same scale $ \bar\Lambda^2$ as in the
double soft limit, so that now
\begin{equation}\label{eq:lamdsnm}
\mathcal{L}= \ln(\N\M).
\end{equation}
In this case, the resummed expression has the same
form as in the double soft case, Eq.~(\ref{RES.SIDIS.DS.1Int.NNLL}), except that the functions
$g_0$ and $D$ are now $M$ dependent; we will denote them by  $\bar{g}_{0,qq}^{\text{xss}}(M)$ and $\bar{D}^{\text{xss}}_{qq}(M)$. We get
\begin{align}
    &A^{(1)}_{q}=C_F\,, \label{eq:xssa1}\\ \label{eq:xssa2}
    &A^{(2)}_{q}=\frac{1}{2}C_F\left[\left(\frac{67}{18}-\frac{\pi ^2}{6}\right) C_{A}-\frac{5}{9} N_{f}\right]\,, \\ \label{eq:xssd1}
    &\bar{D}^{\text{xss},(1)}_{qq}(M)=\bar{\gamma}^{(0)}_{qq}(M)\,, \\ \label{eq:xssd2}
    &\bar{D}^{\text{xss},(2)}_{qq}(M)=D_{qq}^{\text{ds},(2)}-\pi\beta_0
 \bar{ F}(M)+\bar{\gamma}^{(1),t}_{qq,\text{NS}}(M), \\
  \label{eq:xssg1a}
    &\bar{g}_{0,qq}^{\text{xss},(1)}(M)=g_{0,qq}^{\text{ds},(1)}+\bar{F}(M)
    \\
  \label{eq:xssg1b}
  &\phantom{\bar{g}_{0,qq}^{\text{xss},(1)}(M)} = f_2^{(1)}(M)\ln^2{\M}-f_1^{(1)}(M)\ln{\M}+f_0^{(1)}(M)\,, \\
  \label{eq:xssg2}
    &\bar{g}_{0,qq}^{\text{xss},(2)}(M)= f_4^{(2)}(M)\ln^4{\M}-f_3^{(2)}(M)\ln^3{\M}+f_2^{(2)}(M)\ln^2{\M} \nonumber \\
    &\phantom{g_{0,qq}^{(2)}(M)=}-f_1^{(2)}(M)\ln{\M}+f_0^{(2)}(M)\,, 
\end{align}
where the expressions Eq.~(\ref{eq:xssg1b}-\ref{eq:xssg2}) of
$\bar{g}_{0,qq}^{\text{xss},(i)}(M)$ can be obtained by noting that
with the choice of scale  $ \Lambda_{\rm ss}^2$
Eq.~(\ref{eq:sss}), the coefficient of $\ln^kN$ coincides with the
corresponding coefficient  $f_i^{(j)}(M)$ Eq.~(\ref{CTQ2Q_singleN}) of
the fixed-order result, and the
resummed expression with  scale $ \bar\Lambda^2$ Eq.~(\ref{eq:dss}) is obtained by that
with scale  $ \Lambda_{\rm ss}^2$ Eq.~(\ref{eq:sss}) by simply
letting everywhere $\ln^k\N\to(\ln \N+\ln\M)^k$, expanding the result in
powers of $\ln \N$, and then equating order by order
the coefficient of $\ln^k\N$ to  $f_i^{(j)}(M)$.
In Eqs.~(\ref{eq:xssd1}-\ref{eq:xssg1a}) we have defined
\begin{align}\label{eq:fsub}
    \bar{F}(M)= F(M)+C_F\left[\lnM \left(\lnM+\frac{1}{2 M(M+1)}-S_1(M)\right)-\frac{1}{2}  \ln^2\M\right],
     \end{align}    
and  $\bar{\gamma}_{qq}^{(i),t}(M)$ are the anomalous dimensions but
now with 
also all logarithmic contributions subtracted, see Appendix~\ref{spNLO}, so that
\begin{equation}\label{eq:pmtoinf}
  \lim_{M\to\infty} \bar{\gamma}^{(i),t}_{qq}(M)=0.
\end{equation}
Furthermore,     
\begin{equation}\label{eq:ftoinf}
  \lim_{M\to\infty} 
  \bar{F}(M)=0, \end{equation}
because the extra terms in Eq.~(\ref{eq:fsub}) remove
all logarithmic and constant terms from $F(M)$.

Using this form of the resummed expression in the soft limit it is
easier to check that when $M\to\infty$ the double soft result is
reproduced. Indeed, with this choice of scale the resummed expressions
in the double  and single soft limits have the same form, hence the
single soft limit reduces to the double soft limit provide only all
single soft coefficients become identical to their double soft form
when both variables tend to to infinity. It is easy to check that this
is the case:
the coefficients $A^{(i)}$ Eqs.~(\ref{eq:xssa1}-\ref{eq:xssa2}) are the
usual coefficients of the cusp anomalous dimension,  and because of 
Eqs.~(\ref{eq:pmtoinf}-\ref{eq:ftoinf}) the functions $D^{(i)}(M)$ and
$g^{(1)}(M)$ Eqs.~(\ref{eq:xssd1}-\ref{eq:zssh1a}) reduce to the
$M$-independent form that they have in the double soft limit. We have
finally checked explicitly that the constant also reduces to its
double soft expression Eq.~(\ref{eq:g2ds}) in the limit:
\begin{equation}\label{eq:g2toinf}
  \lim_{M\to\infty} \bar{g}_{0,qq}^{\text{xss},(2)}(M) =
  g_{0,qq}^{\text{ds},(2)}.
 \end{equation}

\subsubsection{$z$-single soft}
We turn  to the case in which $M\to\infty$, so the large log is now
\begin{equation}\label{eq:lamdsz}
\mathcal{L}= \ln\M.
\end{equation}
Results can be derived proceeding as in the $x$-single soft case. We now get
\begin{align}
    &A^{(1)}_{q}=C_F\,,\\
    &A^{(2)}_{q}=\frac{1}{2}C_F\left[\left(\frac{67}{18}-\frac{\pi ^2}{6}\right) C_{A}-\frac{5}{9} N_{f}\right]\,, \\ \label{eq:zssd1s}
    &D^{\text{zss},(1)}_{qq}(N)=\Tilde{\gamma}^{(0)}_{qq}(N)\,, \\ \label{eq:zssd2s}
    &D^{\text{zss},(2)}_{qq}(N)=D_{qq}^{\text{ds},(2)}-\pi\beta_0 H(N)+\Tilde{\gamma}^{(1)}_{qq}(N)\,, \\ 
    &g_{0,qq}^{\text{zss},(1)}(N)=g_{0,qq}^{\text{ds},(1)}+ H(N) = h_0^{(1)}(N)\,,\\ 
    &g_{0,qq}^{\text{zss},(2)}(N)=h^{(2)}_0(N),
\end{align}
where the function $H(N)$ is given by
\begin{align}
    H(N) = \frac{1}{2}C_F\left[S^2_1(N)-\frac{S_1(N)}{N(N+1)}+\frac{2 N+1}{N^2 (N+1)}+\zeta_2-S_2(N)\right].
\end{align}

Using instead the double-soft scale Eq.~(\ref{eq:lamdsnm}) we get 
\begin{align}
    &A^{(1)}_{q}=C_F\,,  \label{eq:zssa1}\\ \label{eq:zssa2}
    &A^{(2)}_{q}=\frac{1}{2}C_F\left[\left(\frac{67}{18}-\frac{\pi ^2}{6}\right) C_{A}-\frac{5}{9} N_{f}\right]\,, \\ \label{eq:zssd1}
    &\bar{D}^{\text{zss},(1)}_{qq}(N) =\bar{\gamma}_{qq}^{(0)}(N), \\ \label{eq:zssd2}
    &\bar{D}^{\text{zss},(2)}_{qq}(N) =D_{qq}^{\text{ds},(2)}-\pi \beta_0
  \bar{H}(N) +\bar{\gamma}_{qq,{\rm NS}}^{(1)}(N), \\ \label{eq:zssh1a}
    &\bar{g}_{0,qq}^{\text{zss},(1)}(N)=g_{0,qq}^{\text{ds},(1)}+
  \bar{H}(N) =  h_2^{(1)}(N)\ln^2{\N}-h_1^{(1)}(N)\ln{\N}+h_0^{(1)}(N)\,, \\ \label{eq:zssh2}
    &g_{0,qq}^{\text{zss},(2)}(M)= h_4^{(2)}(N)\ln^4{\N}-h_3^{(2)}(N)\ln^3{\N}+h_2^{(2)}(N)\ln^2{\N} \nonumber \\ 
    &\phantom{C_{c,qq}^{(2)}(M)=}-h_1^{(2)}(N)\ln{\N}+h_0^{(2)}(N),
\end{align}
where now
\begin{align}
  \bar{H}(N) = H(N)+C_F\left[ \lnN \left(\ln\N+\frac{1}{2 N(N+1)}
    -S_1(N)\right)-\frac{1}{2}
    \ln^2\N\right],
\end{align}
$\gamma^{(i)}$ are the usual (spacelike) anomalous dimensions, $\tilde \gamma^{(i)}$
 denotes the anomalous dimensions splitting function with constants
 subtracted and  $\bar{\gamma}^{(i)}$ is the splitting function with both constants and
logs subtracted (see Appendix~\ref{spNLO}) , which satisfies
\begin{equation}\label{eq:pntoinf}
  \lim_{M\to\infty} \bar{\gamma}^{(i)}_{qq}(M)=0.
\end{equation}
Furthermore,     
\begin{equation}\label{eq:htoinf}
  \lim_{M\to\infty} 
  \bar{H}(M)=0. \end{equation}

\subsubsection{Discussion}
The interpretation of the resummed expressions in the double soft and
in the two single soft limits is transparent, and can be summarized
as follows:
\begin{itemize}
\item In the double soft limit, the resummed result is the same as in
  the inclusive case, and consequently the origin and meaning of the
  various contributions to the resummed expression is the same. In
  particular, as discussed in Sect.~\ref{sec:softlimitsgr}, in this
  limit all radiation is soft. The $A$ term then contains
  contributions from radiation that is both soft and collinear, and
  indeed it is  proportional to
  the cusp anomalous dimension, which is by definition the soft limit
  of the standard collinear anomalous dimension. The $D$ contribution
  only starts at NNLL, and it
  collects further soft but non-collinear contributions.
\item Because the double soft limit is a special case of the single
  soft,  if expressed using the same scale Eq.~(\ref{eq:lamdsnm}), the
  single soft result reduces 
  to the double soft  by simply evaluating all coefficients in the
  limit in which the non-soft variable goes to infinity.
\item The nontrivial single-soft generalization of the double-soft
  result is embodied in the coefficients $D^{\text{xss},(i)}_{qq}(M)$
  or $\bar{D}^{\text{xss},(i)}_{qq}(M)$ and $D^{\text{zss},(i)}_{qq}(N)$
  or $\bar{D}^{\text{zss},(i)}_{qq}(N)$. Their meaning is readily
  understood recalling that, as  discussed in
    Sect.~\ref{sec:softlimitsgr}, the purely single soft radiation is
    collinear. Focussing on the case in which  the same scale
    Eq.~(\ref{eq:lamdsnm}) as in the double soft case is adopted, so
    $\bar{D}^{\text{xss},(i)}_{qq}(M)$ and
    $\bar{D}^{\text{zss},(i)}_{qq}(N)$ only contain terms that are
    power suppresses when their respective argument goes to infinity,
    one notes that  the NLL
    coefficient $\bar{D}^{\text{ss},(1)}_{qq}$ is equal to the
    subtracted leading-order anomalous dimensions
    $\bar{\gamma}^{(1)}_{qq}$. In other words, it contains all radiation
    that is collinear to the non-soft variable, up to the soft scale
    given by the soft variable, in agreement with the results of
    Sect.~\ref{sec:softlimitsgr} (recall Eqs.~(\ref{eq:x1cond}-\ref{eq:z1cond}), on top of the soft-collinear already contained in the
    double-soft terms. At  NNLL, the coefficients
    $\bar{D}^{\text{xss},(2)}_{qq}(M)$ and $\bar{D}^{\text{zss},(2)}_{qq}(N)$, which are
    already nonzero in the double-soft limit, on top of a contribution
    equal to the NLO subtracted anomalous dimensions,  $\bar{\gamma}^{(2)}_{qq}(N)$
    or  $\bar{\gamma}_{qq}^{(2),t}(M)$, also contain a contribution
    proportional to the functions $\tilde F(M)$ or  $\tilde
    H(N)$. These (see Eqs.~(\ref{eq:xssg1a},\ref{eq:zssh1a}) correspond
    to the contribution to the NLO constant
    $\bar{g}_{0,qq}^{\text{ss},(1)}(M)$ or respectively
    $\bar{h}_{0,qq}^{\text{ss},(1)}(N)$, namely the $O(\alpha_s)$ contribution to the coefficient function which is
    respectively proportional to $\delta(1-x)$ or $\delta(1-z)$,
    on top of what is already
    present in the double soft limit. Their contribution to the
    $\bar{D}^{\text{xss},(2)}_{qq}(M)$ and
    $\bar{D}^{\text{zss},(2)}_{qq}(N)$ coefficients amounts to running the argument of
    $\alpha_s$ from the hard scale $Q^2$ to the soft scale $Q^2/N$ or
    respectively $Q^2/M$.
    \item The $x$-single soft and $z$-single soft resummation exponents
      coincide up to NLL, because the timelike and spacelike splitting
      functions coincide at leading order
\begin{equation}\label{eq:equalco}
D^{\text{xss},(1)}_{qq}(N)=D^{\text{zss},(1)}_{qq}(N).
\end{equation}
However, this is no longer the case at NLO, and
both the  NLO and NNLO constants and the NNLO
      resummation coefficient found in the two single soft limits differ:
      $g_{0,qq}^{\text{xss},(i)}(N)\not=g_{0,qq}^{\text{zss},(i)}(N)$
      and $D^{\text{xss},(1)}_{qq}(N)\not =D^{\text{zss},(1)}_{qq}(N)$. 
\item As mentioned in the introduction, NLP corrections to the double
  soft
  resummation were determined in Ref.~\cite{Abele:2022wuy}, using
  arguments from Refs.~\cite{Kramer:1996iq,Catani:2001ic}. These were
  found by adding to the double soft resummation coefficient
  $D_{qq}^{\text{ds},(1)}$ Eq.~(\ref{eq:ddsn}) a contribution equal to
  $Q^{(1)}\left(\frac{1}{N}+\frac{1}{M}\right)$, with the coefficient
  $Q^{(1)}$ determined by expanding the  LO anomalous dimension
  $\gamma^{(0)}_{qq}(N)$ as
    \begin{equation}\label{eq:nlp}
    \gamma^{(0)}_{qq}(N)  =-A^{(1)}\ln N+ \gamma^{(0)}_{q,\,\delta}+
      \frac{Q^{(1)}}{N}+\mathcal{O}\left(\frac{1}{N^2}\right).
    \end{equation}
    Because of Eqs.~(\ref{eq:pntoinf},\ref{eq:pmtoinf}), this manifestly corresponds to including the first order
    contribution in the expansion of the single soft coefficients
    $\bar{D}^{\text{xss},(1)}_{qq}(M)$  Eq.~(\ref{eq:xssd1}) and
     $\bar{D}^{\text{zss},(1)}_{qq}(N)$  Eq.~(\ref{eq:zssd1}) in
    powers of of their respective argument about zero. Therefore, our
    results confirm the NLP results of  Ref.~\cite{Abele:2022wuy}, and
    it extends them to the full single soft limit.
\end{itemize}

\section{Conclusions}
We have presented an extension to SIDIS of the asymmetric threshold
resummation formalism, previously developed for the Drell-Yan rapidity
distribution, using SCET~\cite{Lustermans:2019cau,Mistlberger:2025lee},
and  by some of 
us in direct QCD~\cite{de_ros}. This asymmetric resummation
of  processes with two scaling
variables
resums large logs arising when only one of the two variables
is close to threshold. This situation appears to be especially
relevant in the case of SIDIS, where the two scaling variables -- the
Bjorken variable and the fragmentation variable -- play different
physical roles. Our results confirm in this somewhat more general and
perhaps nontrivial
setting the crossing relation between threshold resummation of SIDIS and
of the Drell-Yan rapidity distribution first suggested  in
Ref.~\cite{Sterman:2006hu}. Also, by verifying that the fixed NNLO SIDIS
results agree with the prediction from NLL resummation,
they provide a further nontrivial
check of the correctness of the asymmetric resummation of
Refs.~\cite{Lustermans:2019cau,Mistlberger:2025lee,de_ros}.

These results are  one step further
 in the direction of enabling precision physics
studies at the future EIC, by bringing the theory of deep-inelastic
scattering and related topics up to the modern standards of precision
QCD. They can be used both for precision EIC phenomenology in
kinematic regions in which
threshold resummation is needed in order to achieve a good accuracy, and also in
order to construct an approximate form of higher fixed-order
corrections, as already done in Ref.~\cite{Abele:2022wuy} exploiting
information from the double soft limit. Both results can be arrived at
by combining information from the double soft and the two single soft
limits.

This work provides a first proof of concept towards these goals. In
order to achieve them it will be necessary to extend our results, that
so far are restricted to the nonsinglet quark channel, to include all
partonic channels. For instance, at large $z$ and not too high values
of $Q^2$, small-$x$  initial-state partons provide a dominant
contribution (see e.g.~\cite{Bonino:2025qta}) , so it is crucial to include the singlet contribution to
the non-soft leg in the resummed expression
Eq.~(\ref{RES.SIDIS.SS.1Int.NNLL}). This entails some technical
complications, because the resummation in
Eq.~(\ref{RES.SIDIS.SS.1Int.NNLL}), as mentioned, must be performed in
terms of evolution eigenstates, which, as well-known, are not the
same at different perturbative orders. While the way to handle this
situation is also well-known~\cite{Vogt:2004ns}, its implementation in
a resummed framework requires some care.
Work in this direction is currently ongoing, towards the construction of fully matched
resummed and fixed order results and the study of their
phenomenological implications, in particular at the EIC~\cite{SIDISph}.

\section*{Acknowledgments}
We thank Giovanni Stagnitto for discussions and
help with the results of
Refs.~\cite{Bonino:2024qbh,Bonino:2025qta}. SF thanks Werner Vogelsang
and especially Johannes Michel
for several illuminating discussions.\\
The work of SF and GR has received funding from the European Union
NextGeneration EU program – NRP Mission 4 Component 2 Investment 1.1 –
MUR PRIN 2022 – CUP G53D23001100006 through the Italian Ministry of
University and Research (MUR). The work of FV was partially supported by ERC Advanced Grant 101095857 "Conformal-EIC". SF thanks for hospitality the CERN department of
theoretical physics where part of this work was done.

\appendix

\section{Appendix}
\label{sec:app}

We collect for completeness and in order to fix notation and
conventions the expressions for the running coupling up to NNLO
and of the timelike and spacelike splitting functions up to NLO and of
their Mellin transforms (see e.g. Ref.~\cite{Ellis:1996mzs}). We then
list the coefficients of the 
expansion of the nonsinglet contribution to the Mellin-space coefficient function
$C^T_{qq}(N,M)$ in powers of either of the large  logs
corresponding to the two single soft limits, following the notation of
Eq.~(\ref{CTQ2Q_singleN}) up to NNLO, obtained using the results of Refs.~\cite{Bonino:2024qbh,Bonino:2025qta}.

\subsection{The QCD running coupling}\label{running_coupling}
The running coupling of QCD in the $\overline{\rm MS}$ scheme to NNLL accuracy is given by
\begin{align}
    \alpha_{s}(\mu)
	&=\frac{\alpha_{s}(\mu_{R}) }{\ell}\left[1
		-\frac{\alpha_{s}(\mu_{R}) }{\ell} b_1 \ln\ell\right.\nonumber\\[2mm]
	&+\left.\left(\frac{\alpha_{s}(\mu_{R})}{\ell}\right)^{2} \left(b_1^2 \left(\ln ^2\ell-\ln\ell+l-1\right)-b_2 (\ell-1)\right)\right]
	+\mathcal{O}\left(\alpha_s^3\left(\alpha_s\ln\frac{\mu^2}{\mu_R^2}\right)\right)
\label{eq:asmu}
\end{align}
where
\begin{align}
&\ell=1 +\beta_0 \alpha_{s} (\mu_{R}) \ln \frac{\mu^{2}}{\mu^{2}_{R}}
\\
&\beta_0  =  \frac{1}{12\pi} \left(11C_A-2 N_f\right)
\\
&\beta_1 = \frac{1}{24\pi^2}\left(17C_A^2-5C_AN_f-3C_F N_f\right)
\\
&\beta_2  =  \frac{1}{64\pi^3}\left(\frac{2857}{54} C_A^3- \frac{1415}{54} C_A^2 N_f-\frac{205}{18} C_A C_F N_f+C_F^2 N_f+
    \frac{79}{54} C_A N_f^2+\frac{11}{9} C_F N_f^2\right),   
\end{align}
and $b_1=\frac{\beta_1}{\beta_0},b_2=\frac{\beta_2}{\beta_0}$.
$N_f$ is the number of light flavors and 
\begin{equation}\label{cqcg} 
    C_F\,=\,\frac{N_c^2-1}{2N_c}=\frac{4}{3};\qquad C_A\,=\,N_c=3.    
\end{equation}

\subsection{NLO splitting functions and anomalous dimensions}\label{spNLO}
We provide expressions for the nonsinglet quark-quark splitting function; 
all expressions given here (specifically at NLO)
refer to the pure nonsinglet contribution. We define
\begin{equation}
    P_{qq}=\frac{\as}{\pi}P_{qq}^{(0)}+\left(\frac{\as}{\pi}\right)^2P_{qq}^{(1)}+\mathcal{O}(\as^3).
\end{equation}

The LO time-like and space-like expressions concide and are
given by
\begin{equation}
    P_{qq}^{(0)}(x)=\Tilde P_{qq}^{(0)}(x)+\frac{C_F}{2}\frac{3}{2}\delta(1-x)\,,
\end{equation}
where
\begin{equation}
    \Tilde P_{qq}^{(0)}(x)=\frac{C_F}{2}\frac{1+x^2}{(1-x)_+}\,.
\end{equation}
Their  Mellin transforms are respectively given by
\begin{equation}
    \gamma_{qq}^{(0)}(N)=\Tilde  \gamma_{qq}^{(0)}(N)+ \frac{3}{2}  \frac{C_F}{2},
\end{equation}
and
\begin{equation}
 \Tilde  \gamma_{qq}^{(0)}(N)=  \frac{C_F}{2}\left[\frac{1}{N(N+1)}-2S_1(N)\right].
\end{equation}
The subtracted anomalous dimension which enters the NLO single soft coefficients
Eqs.~(\ref{eq:xssd1},\ref{eq:zssd1}) is finally given by
\begin{equation}
    \bar{\gamma}_{qq}^{(0)}(N)=\frac{C_F}{2} \left[2\lnN+\frac{1}{N(N+1)}-2S_1(N)\right].
\end{equation}

The NLO coefficient of the space-like splitting function  is
\begin{align}
    P_{qq,\text{NS}}^{(1)}(x)&=\frac{1}{4} \left\{
    C_F^2 \left[\left(-2 \ln x \ln (1-x)-\frac{3 \ln x}{2}\right) \left(2 \left[\frac{1}{1-x}\right]_+-x-1\right)\right. \right.
\nonumber\\
    &\qquad\left.-\frac{1}{2} (x+1) \ln^2x -\left(\frac{7 x}{2}+\frac{3}{2}\right) \ln x-5 (1-x)\right]
\nonumber \\
    &+C_F C_A \left[\left(\frac{1}{2} \ln^2x+\frac{11 \ln x}{6}-\frac{\pi ^2}{6}+\frac{67}{18}\right) \left(2 \left[\frac{1}{1-x}\right]_+-x-1\right) \right. 
\nonumber \\
    &\qquad\left.+(x+1) \ln x+\frac{20 (1-x)}{3}\right]
\nonumber \\
    &+\frac{1}{2} C_F N_f \left[\left(-\frac{2}{3} \ln x-\frac{10}{9}\right) \left(2 \left[\frac{1}{1-x}\right]_+-x-1\right)-\frac{4(1-x)}{3}\right]
\nonumber \\
    &+\frac{1}{4} \delta (1-x) 
    \left.\left[C_F^2 \left(6 \zeta_3+\frac{3}{8}-\frac{\pi ^2}{2}\right)
         -C_F C_A \left(-3\zeta_3+\frac{17}{24}+\frac{11 \pi^2}{18}\right)
         -\frac{1}{2}C_F N_f \left(\frac{1}{6}+\frac{2 \pi ^2}{9}\right) 
\right]\right\}.
\end{align}
Its Mellin transform is given by
\begin{align}
    \gamma_{qq,\text{NS}}^{(1)}(N)
    &= \Tilde \gamma_{qq,\text{NS}}^{(1)}(N)
    +\frac{1}{4} \left(C_F^2 \left(6 \zeta_3+\frac{3}{8}-\frac{\pi
    ^2}{2}\right) -C_F C_A \left(-3 \zeta_3+\frac{17}{24}+\frac{11 \pi ^2}{18}\right)-\frac{1}{2} \left(\frac{1}{6}+\frac{2 \pi ^2}{9}\right) C_F N_F\right)\,,
\end{align}
where 
\begin{align}
  \Tilde  \gamma_{qq,\text{NS}}^{(1)}(N) &= C_F^2 \left(-\frac{(2 N+1) S_1(N)}{2 N^2 (N+1)^2}-\frac{\left(3 N^2+3 N+2\right) S_2(N)}{4 N (N+1)}+\frac{\left(3 N^2+3 N+2\right) \zeta_2}{4 N (N+1)} \right. \nonumber \\[2mm]
    &\left.+\frac{3 N^3+N^2-1}{4 N^3 (N+1)^3}-\zeta_2 S_1(N)+S_1(N) S_2(N)+S_3(N)-\zeta_3\right) \nonumber \\[2mm]
    &+C_F N_F \left(-\frac{11 N^2+5 N-3}{36 N^2 (N+1)^2}+\frac{5 S_1(N)}{18}-\frac{S_2(N)}{6}+\frac{\zeta_2}{6}\right) \nonumber \\[2mm]
    &+C_F C_A \left(-\frac{\left(11 N^2+11 N+3\right) \zeta_2}{12 N (N+1)}+\frac{151 N^4+236 N^3+88 N^2+3 N+18}{72 N^3 (N+1)^3} \right. \nonumber \\[2mm]
    &\left.+\frac{1}{2} \zeta_2 S_1(N)-\frac{67 S_1(N)}{36}+\frac{11
    S_2(N)}{12}-\frac{S_3(N)}{2}+\frac{\zeta_3}{2}\right),
\end{align}
and $S_i(N)$ are harmonic sums~\cite{Vermaseren:1998uu}.

The NLO timelike splitting function is given by 
\begin{align}
    P_{qq,\text{NS}}^{(1),t}(x)=P_{qq,\text{NS}}^{(1)}(x)+\Delta_{qq,\text{NS}}^{(1)}(x) \,,
\end{align}
where
\begin{align}
    \Delta_{qq,\text{NS}}^{(1)}(x) &=\frac{1}{16}
    C_F^2 \left(\left(-32 \left[\frac{1}{1-x}\right]_+ +16 x+16\right)
    H_{1,0}(x)+\left(-16 \left[\frac{1}{1-x}\right]_++12
    x+12\right) \ln^2 x \right. \nonumber \\[2mm]
    &\left. +\left(-32 \left[\frac{1}{1-x}\right]_++16
    x+16\right) H_2(x)  + \left(24 \left[\frac{1}{1-x}\right]_+-4
    x-20\right) \ln x \right)\,.
\end{align}
Here  $H_n$ and $H_{n,m}$ are harmonic polylogarithms~\cite{Remiddi:1999ew}. The
Mellin transform of the difference is
\begin{align}
    \Delta_{qq,\text{NS}}^{(1)}(N) &= C_F^2 \Bigg[\frac{(2 N+1) S_1(N)}{N^2 (N+1)^2}+\frac{\left(3 N^2+3 N+2\right) S_2(N)}{2 N (N+1)}-\frac{\left(3 N^2+3 N+2\right) \zeta_2}{2 N (N+1)}  \nonumber \\
    & -\frac{6 N^3+9 N^2+7 N+2}{4 N^3 (N+1)^3}+2 \zeta_2 S_1(N)-2 S_1(N) S_2(N)\Bigg].
\end{align}

The explicit expressions of the spacelike and timelike Mellin-space subtracted
anomalous dimensions that enter the NNLO single soft
coefficients Eqs.~(\ref{eq:xssd2},\ref{eq:zssd2}) are
\begin{align}
    &\bar{\gamma}_{qq,\text{NS}}^{(1)}(N)= C_F^2 \left(-\frac{(2 N+1) S_1(N)}{2 N^2 (N+1)^2}-\frac{\left(3 N^2+3 N+2\right) S_2(N)}{4 N (N+1)}+\frac{\left(3 N^2+3 N+2\right) \zeta_2}{4 N (N+1)} \right.\nonumber \\[2mm]
    &\phantom{\bar{\gamma}_{qq,NS}^{(1)}(N)=}\left. +\frac{3 N^3+N^2-1}{4 N^3 (N+1)^3}-\zeta_2 S_1(N)+S_1(N) S_2(N)+S_3(N)-\zeta_3\right) \nonumber \\[2mm]
    &\phantom{\bar{\gamma}_{qq,NS}^{(1)}(N)=}+C_F N_F \left(-\frac{5 \lnN}{18}-\frac{11 N^2+5 N-3}{36 N^2 (N+1)^2}+\frac{5 S_1(N)}{18}-\frac{S_2(N)}{6}+\frac{\zeta_2}{6}\right) \nonumber \\[2mm]
    &\phantom{\bar{\gamma}_{qq,NS}^{(1)}(N)=}+C_F C_A \left(-\frac{1}{2} \zeta_2 \lnN+\frac{67 \lnN}{36}-\frac{\left(11 N^2+11 N+3\right) \zeta_2}{12 N (N+1)} \right. \nonumber \\[2mm]
    &\phantom{\bar{\gamma}_{qq,NS}^{(1)}(N)=}+\frac{151 N^4+236 N^3+88 N^2+3 N+18}{72 N^3 (N+1)^3} \nonumber \\[2mm]
    &\phantom{\bar{\gamma}_{qq,NS}^{(1)}(N)=}\left.+\frac{1}{2} \zeta_2 S_1(N)-\frac{67 S_1(N)}{36}+\frac{11 S_2(N)}{12}-\frac{S_3(N)}{2}+\frac{\zeta_3}{2}\right)\,,
\end{align}
and 
\begin{align}\label{eq:pqqt}
    & \bar{\gamma}^{(1),t}_{qq,\text{NS}}(N)= C_F^2 \left(\frac{(2 N+1) S_1(N)}{2 N^2 (N+1)^2}+\frac{\left(3N^2+3 N+2\right) S_2(N)}{4 N (N+1)}-\frac{\left(3 N^2+3 N+2\right) \zeta_2}{4 N (N+1)} \right. \nonumber \\[2mm]
    &\phantom{\bar{\gamma}_{qq,NS}^{(1)}(N)=} \left.-\frac{3 N^3+8 N^2+7 N+3}{4 N^3 (N+1)^3}+\zeta_2 S_1(N)-S_1(N) S_2(N)+S_3(N)-\zeta_3\right) \nonumber \\[2mm]
    &\phantom{\bar{\gamma}_{qq,NS}^{(1)}(N)=}+C_F N_F \left(-\frac{5 \lnN}{18}-\frac{11 N^2+5 N-3}{36 N^2 (N+1)^2}+\frac{5 S_1(N)}{18}-\frac{S_2(N)}{6}+\frac{\zeta_2}{6}\right) \nonumber \\[2mm]
    &\phantom{\bar{\gamma}_{qq,NS}^{(1)}(N)=}+C_F C_A \left(-\frac{1}{2} \zeta_2 \lnN+\frac{67 \lnN}{36}-\frac{\left(11 N^2+11 N+3\right) \zeta_2}{12 N (N+1)} \right. \nonumber \\[2mm]
    &\phantom{\bar{\gamma}_{qq,NS}^{(1)}(N)=}+\frac{151 N^4+236 N^3+88 N^2+3 N+18}{72 N^3 (N+1)^3} \nonumber \\[2mm]
    &\phantom{\bar{\gamma}_{qq,NS}^{(1)}(N)=}\left.+\frac{1}{2} \zeta_2 S_1(N)-\frac{67 S_1(N)}{36}+\frac{11 S_2(N)}{12}-\frac{S_3(N)}{2}+\frac{\zeta_3}{2}\right) \,.
\end{align}

\subsection{Coefficient functions in soft limits}
\label{CFSofts}

We list the coefficients that enter Eq.~(\ref{CTQ2Q_singleN}).

\subsubsection{$x$ single soft}\label{app:CFSoftx}

NLO coefficients
\begin{align}
   &f^{(1)}_2(M) = \frac{\Cf}{2} \\[2mm]
    &f^{(1)}_1(M) = \Cf \left(S_1(M)-\frac{1}{2 M^2+2 M}\right) \\[2mm]
    &f^{(1)}_0(M) = \Cf\Bigg(\frac{ 2 M^2-M-1}{2 M^2 (M+1)^2}+\frac{1}{2}  S_1(M){}^2-\frac{ S_1(M)}{2 M (M+1)} 
    +\frac{3}{2}  S_2(M)-\frac{ \zeta_2}{2}-4  \Bigg)
\end{align}

NNLO coefficients
\begin{align}
    &f^{(2)}_4(M) = \frac{1}{8} \Cf^2\,, \\[2mm]
    &f^{(2)}_3(M) = \Cf^2 \left(\frac{S_1(M)}{2}-\frac{1}{4 M (M+1)}\right)+\frac{11 \Cf \Ca}{72}-\frac{\Cf \Nf}{36}\,, \\[2mm]
    &f^{(2)}_2(M) = \Cf^2 \left(\frac{4 M^2-2 M-1}{8 M^2 (M+1)^2}+\frac{3}{4} S_1(M){}^2-\frac{3 S_1(M)}{4 M (M+1)}+\frac{3 S_2(M)}{4}-\frac{\zeta_2}{4}-2\right) \nonumber \\[2mm]
    &\phantom{f^{(2)}_2(M) =}+\Cf \Ca \left(\frac{11 S_1(M)}{24}-\frac{11}{48 M (M+1)}-\frac{\zeta_2}{4}+\frac{67}{72}\right) \nonumber \\[2mm]
    &\phantom{f^{(2)}_2(M) =}+\Cf \Nf \left(-\frac{1}{12} S_1(M)+\frac{1}{24 M (M+1)}-\frac{5}{36}\right)\,,
\end{align}

\begin{align}
    &f^{(2)}_1(M) = \Cf^2 \left(-\frac{\left(3 M^2+3 M+5\right) S_2(M)}{4 M (M+1)}+\frac{3 \left(M^2+M+1\right) \zeta_2}{4 M (M+1)} \right. \nonumber \\[2mm]
    &\phantom{f^{(2)}_1(M) =}\left.-\frac{\left(16 M^4+32 M^3+12 M^2+6 M+3\right) S_1(M)}{4 M^2 (M+1)^2}+\frac{8 M^4+19 M^3+14 M^2+8 M+4}{4 M^3 (M+1)^3} \right. \nonumber \\[2mm]
    &\phantom{f^{(2)}_1(M) =}\left.-\frac{3}{2} \zeta_2 S_1(M)+\frac{1}{2} S_1(M){}^3-\frac{3 S_1(M){}^2}{4 M (M+1)}+\frac{5}{2} S_2(M) S_1(M)-S_3(M)+\zeta_3\right) \nonumber \\[2mm]
    &\phantom{f^{(2)}_1(M) =}+\Cf N_f \left(-\frac{\left(10 M^2+10 M-3\right) S_1(M)}{36 M (M+1)}-\frac{1}{12} S_1(M){}^2-\frac{S_2(M)}{12} \right. \nonumber \\[2mm] 
    &\phantom{f^{(2)}_1(M) =}\left.+\frac{5 M+8}{36 M (M+1)^2} +\frac{\zeta_2}{12}-\frac{7}{27}\right) \nonumber \\[2mm]
    &\phantom{f^{(2)}_1(M) =}+\Cf \Ca \left(\frac{\left(134 M^2+134 M-33\right) S_1(M)}{72 M (M+1)}-\frac{85 M^4+203 M^3+154 M^2+36 M+18}{72 M^3 (M+1)^3} \right. \nonumber \\[2mm] 
    &\phantom{f^{(2)}_1(M) =}\left.-\frac{1}{2} \zeta_2 S_1(M)+\frac{11}{24} S_1(M){}^2+\frac{11 S_2(M)}{24}\right. \nonumber \\[2mm]
    &\phantom{f^{(2)}_1(M) =}\left.+\frac{S_3(M)}{2}+\frac{\zeta_2}{4 M (M+1)}-\frac{11 \zeta_2}{24}-\frac{9 \zeta_3}{4}+\frac{101}{54}\right)\,, 
\end{align}

\begin{align}
    &f^{(2)}_0(M) = \Cf^2\left(\frac{1}{8} S_1(M){}^4-\frac{S_1(M){}^3}{4 M (M+1)}-\frac{5}{4} \zeta_2 S_1(M){}^2 \right. \nonumber \\[2mm]
    &\left.+\frac{\left(-M^5+13 M^4+41 M^3+15 M^2+2\right) S_1(M)}{8 M^3 (M+1)^3}+\frac{5 \zeta_2 S_1(M)}{4 M^2+4 M}-\frac{1}{2} \zeta_3 S_1(M) \right. \nonumber \\[2mm]
    &\left.+\frac{S_1(M)^2 \left(7 M^2 (M+1)^2 S_2(M)-2 \left(4 M^4+8 M^3+3 M^2+2 M+1\right)\right)}{4 M^2 (M+1)^2} \right. \nonumber \\[2mm]
    & +\left. \frac{\left(-7 M (M+1) S_2(M)\right)S_1(M)}{4 M^2 (M+1)^2}+\frac{1}{2} \left(5 S_3(M)-2 S_{2,1}(M)\right) S_1(M)-\frac{31 \zeta_2^2}{40} \right. \nonumber \\[2mm]
    & \left.+\frac{-33 M^6-50 M^5+29 M^4+81 M^3+65 M^2+48 M+15}{8 M^4 (M+1)^4} \right. \nonumber \\[2mm]
    &\left.+\frac{\left(4 M^2-2 M-1\right) \zeta_2}{8 M^2 (M+1)^2}+\frac{\left(19 M^4+38 M^3+12 M^2+M+1\right) \zeta_2}{4 M^2 (M+1)^2}+\frac{13 \zeta_2}{8} \right. \nonumber \\[2mm]
    &-\frac{\zeta_3}{2 M (M+1)} -\frac{15 \zeta_3}{4}\nonumber \\[2mm]
    &\left.+\frac{3 \left(7 \zeta_3 M^2+7 \zeta_3 M+2 \zeta_3\right)}{8 M (M+1)}-\frac{\left(27 M^4+54 M^3+18 M^2+7 M+5\right) S_2(M)}{4 M^2 (M+1)^2} \right. \nonumber \\[2mm]
    & \left.-\frac{5}{4} \zeta_2 S_2(M)-\frac{\left(9 M^2+9 M+10\right) S_3(M)}{8 M (M+1)}-\frac{33 S_4(M)}{8}-\frac{\left(3 M^2+3 M-2\right) S_{2,1}(M)}{4 M (M+1)} \right.\nonumber \\[2mm]
    &\left.+\frac{13}{4} S_{2,2}(M)-2 S_{3,1}(M)+\frac{3}{2} S_{2,1,1}(M)+\frac{511}{64}\right) \nonumber \\[2mm]
    &+\frac{1}{36} (11 \Ca-2 N_f) \zeta_3 \Cf \nonumber \\[2mm]
    & +N_f\Cf \left(-\frac{1}{36} S_1(M){}^3-\frac{\left(10 M^2+10 M-3\right) S_1(M){}^2}{72 M (M+1)} \right. \nonumber\\[2mm]
    & \left.-\frac{\left(28 M^3+56 M^2+13 M-24\right) S_1(M)}{108 M (M+1)^2}+\frac{1}{12} \zeta_2 S_1(M)-\frac{1}{12} S_2(M) S_1(M) \right. \nonumber \\[2mm]
    &\left.-\frac{11 M^4-74 M^3-109 M^2-6 M+9}{216 M^3 (M+1)^3}+\frac{\left(5 M^2+5 M-3\right) \zeta_2}{36 M (M+1)} \right. \nonumber\\[2mm]
    &\phantom{f^{(2)}_0(M) =}\left.+\frac{\zeta_2}{24 M (M+1)}+\frac{7 \zeta_2}{18}+\frac{\zeta_3}{12}-\frac{\left(20 M^2+20 M-3\right) S_2(M)}{72 M (M+1)}+\frac{S_3(M)}{36}+\frac{127}{96}\right) \nonumber \\[2mm]
    &+\Ca\Cf \left(\frac{11}{72} S_1(M){}^3+\frac{\left(404 M^5+1239 M^4+1038 M^3-70 M^2-273 M+54\right) S_1(M)}{216 M^2 (M+1)^3}\right. \nonumber \\[2mm]
    &\left.-\frac{11}{24} \zeta_2 S_1(M)-\frac{13}{4} \zeta_3 S_1(M) \right. \nonumber \\[2mm]
    &\left.-\frac{S_1(M)^2 \left(-134 M^2+36 (M+1) S_2(M) M-134 M+33\right)}{144 M (M+1)} \right. \displaybreak[1] \nonumber \\[2mm]
    &\left. \frac{-6 S_1(M) \left(11 M^2+11 M+6\right) S_2(M)}{144 M (M+1)}-\frac{1}{2} \left(S_3(M)-2 S_{2,1}(M)\right) S_1(M)+\frac{21 \zeta_2^2}{20}\right. \nonumber\\[2mm]
    &\left.+\frac{124 M^6-1044 M^5-2679 M^4-2396 M^3-1002 M^2-765 M-270}{432 M^4 (M+1)^4} \right. \nonumber \\[2mm]
    &\left.-\frac{11 \zeta_2}{48 M (M+1)}-\frac{101 \zeta_2}{36}-\frac{29 \zeta_2 M^4+58 \zeta_2 M^3+17 \zeta_2 M^2-30 \zeta_2 M-9 \zeta_2}{36 M^2 (M+1)^2} \right. \nonumber \\[2mm]
    & \left. +\frac{\left(11 M^2+11 M+39\right) \zeta_3}{24 M (M+1)}+\frac{43 \zeta_3}{12}+\frac{\left(250 M^3+250 M^2-69 M-36\right) S_2(M)}{144 M^2 (M+1)} \right. \nonumber \\[2mm]
    &\left.-\frac{1}{2} \zeta_2 S_2(M)-\frac{\left(11 M^2+11 M-18\right) S_3(M)}{72 M (M+1)}+S_4(M)-\frac{S_{2,1}(M)}{2 M (M+1)} \right. \nonumber\\[2mm]
    & \left.+S_{2,2}(M)+S_{3,1}(M)-\frac{3}{2} S_{2,1,1}(M)-\frac{1535}{192}\right) 
\end{align}

\subsubsection{$z$ single soft}\label{app:CFSoftz}

NLO coefficients
\begin{align}
    &h_2^{(1)}(N) = \frac{C_F}{2} \\[2mm]
    &h_1^{(1)}(N) = C_F \left(S_1(N)-\frac{1}{2 N^2+2 N}\right) \\[2mm]
    &h_0^{(1)}(N) = C_F \left(\frac{2 N+1}{2 N^2 (N+1)}+\frac{1}{2} S_1(N){}^2-\frac{S_1(N)}{2 N (N+1)}-\frac{S_2(N)}{2}+\frac{3 \zeta_2}{2}-4\right)
\end{align}

NNLO coefficients
\begin{align}
    &h_4^{(2)}(N) = \frac{1}{8} C_F^2 \\[2mm]
    &h_3^{(2)}(N) = C_F^2 \left(\frac{S_1(N)}{2}-\frac{1}{4 N (N+1)}\right)+\frac{11}{72} C_F C_A -\frac{\Cf N_f}{36}
\end{align}

\begin{align}
    &h_2^{(2)}(N) = C_F^2 \left(\frac{4 N^2+6 N+3}{8 N^2 (N+1)^2}+\frac{3}{4} S_1(N){}^2-\frac{3 S_1(N)}{4 N(N+1)}-\frac{S_2(N)}{4}+\frac{3 \zeta_2}{4}-2\right) \nonumber \\[2mm]
    &\phantom{h_2^{(2)}(N) =}+C_F C_A \left(\frac{11 S_1(N)}{24}-\frac{11}{48 N (N+1)}-\frac{\zeta_2}{4}+\frac{67}{72}\right) \nonumber \\[2mm]
    &\phantom{h_2^{(2)}(N) =} +C_F N_f \left(-\frac{1}{12} S_1(N)+\frac{1}{24 N (N+1)}-\frac{5}{36}\right) 
\end{align}

\begin{align}
    &h_1^{(2)}(N) = C_F^2 \left(\frac{3 \left(N^2+N+1\right) S_2(N)}{4 N (N+1)}-\frac{\left(3 N^2+3 N+5\right) \zeta_2}{4 N (N+1)}+\frac{8 N^3+13 N^2+5 N-3}{4 N^2 (N+1)^3} \right. \nonumber \\[2mm]
    &\phantom{h_1^{(2)}(N) =}\left.-\frac{\left(16 N^4+32 N^3+12 N^2-10 N-5\right) S_1(N)}{4 N^2 (N+1)^2}+\frac{5}{2} \zeta_2 S_1(N)+\frac{1}{2} S_1(N){}^3 \right. \nonumber \\[2mm] 
    &\phantom{h_1^{(2)}(N) =}\left.-\frac{3 S_1(N){}^2}{4 N (N+1)}-\frac{3}{2} S_2(N) S_1(N)-S_3(N)+\zeta_3\right) \nonumber \\[2mm]
    &\phantom{h_1^{(2)}(N) =} +C_F N_f \left(-\frac{\left(10 N^2+10 N-3\right) S_1(N)}{36 N (N+1)}+\frac{5 N^2-4 N-6}{36 N^2 (N+1)^2}\right.\nonumber \\[2mm]
    &\phantom{h_1^{(2)}(N) =}\left.-\frac{1}{12} S_1(N){}^2+\frac{S_2(N)}{4}-\frac{\zeta_2}{4}-\frac{7}{27}\right) \nonumber \\[2mm]
    &\phantom{h_1^{(2)}(N) =}+C_F C_A \left(\frac{\left(134 N^2+134 N-33\right) S_1(N)}{72 N (N+1)}-\frac{85 N^4+71 N^3-44 N^2-30 N+18}{72 N^3 (N+1)^3}\right. \nonumber\\[2mm]
    &\phantom{h_1^{(2)}(N) =}\left.-\frac{1}{2} \zeta_2 S_1(N)+\frac{11}{24} S_1(N){}^2-\frac{11 S_2(N)}{8}+\frac{S_3(N)}{2}+\frac{11 N \zeta_2}{6 (N+1)}+\frac{\zeta_2}{4 N (N+1)} \right. \nonumber \\[2mm]
    &\phantom{h_1^{(2)}(N) =}\left.+\frac{11 \zeta_2}{6 (N+1)}-\frac{11 \zeta_2}{24}-\frac{9 \zeta_3}{4}+\frac{101}{54}\right) 
\end{align}

\begin{align}
    &h_0^{(2)}(N) = C_F^2\left(\frac{1}{8} S_1(N){}^4-\frac{S_1(N){}^3}{4 N (N+1)}+\frac{7}{4} \zeta_2 S_1(N){}^2+\frac{\left(N^4+17 N^3+4 N^2-24 N-8\right) S_1(N)}{8 N^3 (N+1)^2} \right. \nonumber \\[2mm]
    &-\frac{3 \zeta_2 S_1(N)}{4 N (N+1)}-\frac{\zeta_2 S_1(N)}{N^2+N}-\frac{7}{2} \zeta_3 S_1(N)  \nonumber \\[2mm] 
    &-\frac{ \left(8 N^4+16 N^3+5 (N+1)^2 S_2(N) N^2+6 N^2-6 N-3\right) S_1(N)^2}{4 N^2 (N+1)^2} \nonumber \\[2mm]
    &+\frac{5 S_2(N) S_1(N)}{4 N (N+1)}+\frac{3}{2} S_3(N) S_1(N)+S_{2,1}(N) S_1(N) \nonumber \\[2mm] 
    &-\frac{11 \zeta_2^2}{40}+\frac{2 N^7-25 N^6-106 N^5-142 N^4-83 N^3-21 N^2-13 N-5}{8 N^4 (N+1)^4} \nonumber \\[2mm] 
    &+\frac{\left(4 N^2+6 N+3\right) \zeta_2}{8 N^2 (N+1)^2}-\frac{\left(20 N^3+40 N^2+13 N-3\right) \zeta_2}{4 N (N+1)^2}+\frac{13 \zeta_2}{8}-\frac{\zeta_3}{2 N (N+1)}-\frac{15 \zeta_3}{4} \nonumber \\[2mm]
    &-\frac{3 \left(\zeta_3 N^2+\zeta_3 N-6 \zeta_3\right)}{8 N (N+1)}+\frac{\left(12 N^3+24 N^2+7 N-1\right) S_2(N)}{4 N (N+1)^2}+\frac{3}{4} \zeta_2 S_2(N) \nonumber \\[2mm]
    &-\frac{3 \left(3 N^2+3 N+2\right) S_3(N)}{8 N (N+1)}+\frac{23 S_4(N)}{8}+\frac{\left(3 N^2+3 N-2\right) S_{2,1}(N)}{4 N (N+1)}\nonumber \\[2mm]
    &\left.-\frac{1}{4} S_{2,2}(N)-\frac{1}{2} S_{3,1}(N)-\frac{3}{2} S_{2,1,1}(N)+\frac{511}{64}\right)  \nonumber \\[2mm]
    &+\frac{1}{36} (11 C_A-2 N_f) \zeta_3 C_F \nonumber \\[2mm]
    & +C_F\left(\frac{S_1(N){}^2 \left(N^2+N+2\right)^2}{16 N^2 (N+1)^2 \left(N^2+N-2\right)}+\frac{\zeta_2 \left(N^2+N+2\right)^2}{16 N^2 (N+1)^2 \left(N^2+N-2\right)} \right. \nonumber \\[2mm]
    &-\frac{S_2(N) \left(N^2+N+2\right)^2}{16N^2 (N+1)^2 \left(N^2+N-2\right)} \nonumber \\[2mm]
    & +\frac{N^{10}+8 N^9+33 N^8+127 N^7+459 N^6+1111 N^5}{16 (N-1) N^4 (N+1)^4 (N+2)^3} \nonumber \\[2mm]
    & +\frac{1725 N^4+1776 N^3+1192 N^2+464 N+80}{16 (N-1) N^4 (N+1)^4 (N+2)^3} \nonumber \\[2mm]
    &\left. -\frac{\left(N^6+12 N^5+53 N^4+86 N^3+80 N^2+56 N+16\right) S_1(N)}{8 (N-1) N^3 (N+1)^3 (N+2)^2}\right) \nonumber \\[2mm]
    &+N_f C_F \left(-\frac{1}{36} S_1(N){}^3-\frac{\left(10 N^2+10 N-3\right) S_1(N){}^2}{72 N (N+1)}-\frac{1}{12} \zeta_2 S_1(N)+\frac{1}{12} S_2(N) S_1(N) \right. \nonumber \\[2mm]
    &-\frac{7 N^2 S_1(N)}{27 (N+1)^2}-\frac{14 N S_1(N)}{27 (N+1)^2}-\frac{S_1(N)}{9 N (N+1)^2}-\frac{S_1(N)}{6 N^2 (N+1)^2}-\frac{13 S_1(N)}{108 (N+1)^2} \nonumber \displaybreak[1] \\[2mm]
    &-\frac{11 N^4+46 N^3-19 N^2-90 N-45}{216 N^3 (N+1)^3}+\frac{\zeta_2}{24 N (N+1)}-\frac{\zeta_2}{36}+\frac{\zeta_3}{12} \nonumber \\[2mm]
    &\left.+\frac{\left(20 N^2+20 N-3\right) S_2(N)}{72 N (N+1)}-\frac{11 S_3(N)}{36}+\frac{1}{6} S_{2,1}(N)+\frac{127}{96}\right) \nonumber \\[2mm]
    &+C_FC_A \left(\frac{11 N S_1(N){}^3}{72 (N+1)}+\frac{11 S_1(N){}^3}{72 (N+1)}-\frac{3}{4} \zeta_2 S_1(N){}^2 \right. \nonumber \\[2mm]
    & +\frac{\left(404 N^5+781 N^4+122 N^3-12 N^2+144 N-54\right) S_1(N)}{216 N^3 (N+1)^2} \nonumber \\[2mm]
    &+\frac{11 N \zeta_2 S_1(N)}{24 (N+1)}+\frac{3 \zeta_2 S_1(N)}{4 N (N+1)} +\frac{11 \zeta_2 S_1(N)}{24 (N+1)}+\frac{5}{4} \zeta_3 S_1(N) \nonumber \\[2mm]
    &+\frac{\left(S_1(N) \left(134 N^2+72 (N+1) S_2(N) N+134 N-33\right)\right) S_1(N)}{144 N (N+1)} \nonumber \\[2mm]
    &-\frac{\left(6 \left(11 N^2+11 N+12\right) S_2(N)\right) S_1(N)}{144 N (N+1)} \nonumber \\[2mm]
    &-\frac{108 N^7+308 N^6-348 N^5-999 N^4-298 N^3+324 N^2-153 N-162}{432 N^4 (N+1)^4}  \nonumber \\[2mm]
    &-\frac{3}{2} S_{2,1}(N) S_1(N)-\frac{19 \zeta_2^2}{20}+\frac{\left(70 N^4+140 N^3+49 N^2-15 N-6\right) \zeta_2}{24 N^2 (N+1)^2}\nonumber \\[2mm]
    &-\frac{11 \zeta_2}{48 N (N+1)}-\frac{101 \zeta_2}{36}+\frac{\left(11 N^2+11 N-15\right) \zeta_3}{24 N (N+1)}+\frac{43 \zeta_3}{12}\nonumber \\[2mm]
    &-\frac{\left(286 N^3+572 N^2+181 N-33\right) S_2(N)}{144 N (N+1)^2}-\frac{1}{4} \zeta_2 S_2(N)+\frac{121 S_3(N)}{72}-\frac{5 S_4(N)}{4}\nonumber \\[2mm]
    &\left.-\frac{\left(11 N^2+11 N-9\right) S_{2,1}(N)}{12 N (N+1)}-\frac{1}{2} S_{2,2}(N)+2 S_{2,1,1}(N)-\frac{1535}{192}\right) 
\end{align}

\bibliographystyle{UTPstyle}
\bibliography{SIDIS}

@article{Kramer:1996iq,
    author = "Kramer, Michael and Laenen, Eric and Spira, Michael",
    title = "{Soft gluon radiation in Higgs boson production at the LHC}",
    eprint = "hep-ph/9611272",
    archivePrefix = "arXiv",
    reportNumber = "CERN-TH-96-231, DESY-96-170",
    doi = "10.1016/S0550-3213(97)00679-2",
    journal = "Nucl. Phys. B",
    volume = "511",
    pages = "523--549",
    year = "1998"
}

@article{Catani:2001ic,
    author = "Catani, Stefano and de Florian, Daniel and Grazzini, Massimiliano",
    title = "{Higgs production in hadron collisions: Soft and virtual QCD corrections at NNLO}",
    eprint = "hep-ph/0102227",
    archivePrefix = "arXiv",
    reportNumber = "CERN-TH-2001-044",
    doi = "10.1088/1126-6708/2001/05/025",
    journal = "JHEP",
    volume = "05",
    pages = "025",
    year = "2001"
}

@article{Mistlberger:2025lee,
    author = "Mistlberger, Bernhard and Vita, Gherardo",
    title = "{Collinear Approximations for LHC Cross Sections: Factorization and Resummation}",
    eprint = "2502.17548",
    archivePrefix = "arXiv",
    primaryClass = "hep-ph",
    reportNumber = "CERN-TH-2025-041, SLAC-PUB-250203",
    month = "2",
    year = "2025"
}

@article{Forte:2002ni,
    author = "Forte, Stefano and Ridolfi, Giovanni",
    title = "{Renormalization group approach to soft gluon resummation}",
    eprint = "hep-ph/0209154",
    archivePrefix = "arXiv",
    reportNumber = "RM3-TH-02-02, GEF-TH-11-02",
    doi = "10.1016/S0550-3213(02)01034-9",
    journal = "Nucl. Phys. B",
    volume = "650",
    pages = "229--270",
    year = "2003"
}

@article{Forte:2021wxe,
    author = "Forte, Stefano and Ridolfi, Giovanni and Rota, Simone",
    title = "{Threshold resummation of transverse momentum distributions beyond next-to-leading log}",
    eprint = "2106.11321",
    archivePrefix = "arXiv",
    primaryClass = "hep-ph",
    reportNumber = "TIF-UNIMI-2021-08",
    doi = "10.1007/JHEP08(2021)110",
    journal = "JHEP",
    volume = "08",
    pages = "110",
    year = "2021"
}

@Unpublished{SIDISph,
      author         = "Forte, Stefano and Groenendijk, Eva and Ridolfi, Giovanni and Stagnitto, Giovanni",
      note = "{\it In preparation}",
      year           = "2026",
}

@article{Anderle:2012rq,
    author = "Anderle, Daniele P. and Ringer, Felix and Vogelsang, Werner",
    title = "{QCD resummation for semi-inclusive hadron production processes}",
    eprint = "1212.2099",
    archivePrefix = "arXiv",
    primaryClass = "hep-ph",
    doi = "10.1103/PhysRevD.87.034014",
    journal = "Phys. Rev. D",
    volume = "87",
    number = "3",
    pages = "034014",
    year = "2013"
}

@article{Lustermans:2019cau,
    author = "Lustermans, Gillian and Michel, Johannes K. L. and Tackmann, Frank J.",
    title = "{Generalized Threshold Factorization with Full Collinear Dynamics}",
    eprint = "1908.00985",
    archivePrefix = "arXiv",
    primaryClass = "hep-ph",
    reportNumber = "DESY 19-135, DESY-19-135, NIKHEF 2019-021",
    month = "8",
    year = "2019"
}

@article{Abele:2022wuy,
    author = "Abele, Maurizio and de Florian, Daniel and Vogelsang, Werner",
    title = "{Threshold resummation at NLL3 accuracy and approximate N3LO corrections to semi-inclusive DIS}",
    eprint = "2203.07928",
    archivePrefix = "arXiv",
    primaryClass = "hep-ph",
    doi = "10.1103/PhysRevD.106.014015",
    journal = "Phys. Rev. D",
    volume = "106",
    number = "1",
    pages = "014015",
    year = "2022"
}

@article{Abele:2021nyo,
    author = "Abele, Maurizio and de Florian, Daniel and Vogelsang, Werner",
    title = "{Approximate NNLO QCD corrections to semi-inclusive DIS}",
    eprint = "2109.00847",
    archivePrefix = "arXiv",
    primaryClass = "hep-ph",
    doi = "10.1103/PhysRevD.104.094046",
    journal = "Phys. Rev. D",
    volume = "104",
    number = "9",
    pages = "094046",
    year = "2021"
}

@article{Catani:1989ne,
    author = "Catani, S. and Trentadue, L.",
    title = "{Resummation of the QCD Perturbative Series for Hard Processes}",
    reportNumber = "DFF-93/3/89",
    doi = "10.1016/0550-3213(89)90273-3",
    journal = "Nucl. Phys. B",
    volume = "327",
    pages = "323--352",
    year = "1989"
}

@article{Sterman:2006hu,
    author = "Sterman, George F. and Vogelsang, Werner",
    title = "{Crossed Threshold Resummation}",
    eprint = "hep-ph/0606211",
    archivePrefix = "arXiv",
    reportNumber = "BNL-NT-06-20, YITP-SB-06-18",
    doi = "10.1103/PhysRevD.74.114002",
    journal = "Phys. Rev. D",
    volume = "74",
    pages = "114002",
    year = "2006"
}

@article{Bonino:2025qta,
    author = {Bonino, Leonardo and Gehrmann, Thomas and L{\"o}chner, Markus and Sch{\"o}nwald, Kay and Stagnitto, Giovanni},
    title = "{Neutral and charged current semi-inclusive deep-inelastic scattering at NNLO QCD}",
    eprint = "2506.19926",
    archivePrefix = "arXiv",
    primaryClass = "hep-ph",
    reportNumber = "ZU-TH 45/25, INT-PUB-25-018",
    doi = "10.1007/JHEP10(2025)016",
    journal = "JHEP",
    volume = "10",
    pages = "016",
    year = "2025"
}

@article{Vogt:2004ns,
    author = "Vogt, A.",
    title = "{Efficient evolution of unpolarized and polarized parton distributions with QCD-PEGASUS}",
    eprint = "hep-ph/0408244",
    archivePrefix = "arXiv",
    reportNumber = "NIKHEF-04-011",
    doi = "10.1016/j.cpc.2005.03.103",
    journal = "Comput. Phys. Commun.",
    volume = "170",
    pages = "65--92",
    year = "2005"
}

@article{Bonino:2024qbh,
    author = "Bonino, Leonardo and Gehrmann, Thomas and Stagnitto, Giovanni",
    title = "{Semi-Inclusive Deep-Inelastic Scattering at Next-to-Next-to-Leading Order in QCD}",
    eprint = "2401.16281",
    archivePrefix = "arXiv",
    primaryClass = "hep-ph",
    reportNumber = "ZU-TH 07/24",
    doi = "10.1103/PhysRevLett.132.251901",
    journal = "Phys. Rev. Lett.",
    volume = "132",
    number = "25",
    pages = "251901",
    year = "2024"
}

@article{Goyal:2023zdi,
    author = "Goyal, Saurav and Moch, Sven-Olaf and Pathak, Vaibhav and Rana, Narayan and Ravindran, V.",
    title = "{Next-to-Next-to-Leading Order QCD Corrections to Semi-Inclusive Deep-Inelastic Scattering}",
    eprint = "2312.17711",
    archivePrefix = "arXiv",
    primaryClass = "hep-ph",
    doi = "10.1103/PhysRevLett.132.251902",
    journal = "Phys. Rev. Lett.",
    volume = "132",
    number = "25",
    pages = "251902",
    year = "2024"
}

@article{Goyal:2024emo,
    author = "Goyal, Saurav and Lee, Roman N. and Moch, Sven-Olaf and Pathak, Vaibhav and Rana, Narayan and Ravindran, V.",
    title = "{NNLO QCD corrections to unpolarized and polarized SIDIS}",
    eprint = "2412.19309",
    archivePrefix = "arXiv",
    primaryClass = "hep-ph",
    doi = "10.1103/PhysRevD.111.094007",
    journal = "Phys. Rev. D",
    volume = "111",
    number = "9",
    pages = "094007",
    year = "2025"
}

@article{Banerjee:2018vvb,
    author = "Banerjee, Pulak and Das, Goutam and Dhani, Prasanna K. and Ravindran, V.",
    title = "{Threshold resummation of the rapidity distribution for Drell-Yan production at NNLO+NNLL}",
    eprint = "1805.01186",
    archivePrefix = "arXiv",
    primaryClass = "hep-ph",
    reportNumber = "IMSc/2018/05/03, DESY 18-067, IMSC-2018-05-03, DESY-18-067",
    doi = "10.1103/PhysRevD.98.054018",
    journal = "Phys. Rev. D",
    volume = "98",
    number = "5",
    pages = "054018",
    year = "2018"
}

@article{Ravindran:2022aqr,
    author = "Ravindran, V. and Sankar, Aparna and Tiwari, Surabhi",
    title = "{Resummed next-to-soft corrections to rapidity distribution of Higgs boson to NNLO+NNLL\textasciimacron{}}",
    eprint = "2205.11560",
    archivePrefix = "arXiv",
    primaryClass = "hep-ph",
    doi = "10.1103/PhysRevD.108.014012",
    journal = "Phys. Rev. D",
    volume = "108",
    number = "1",
    pages = "014012",
    year = "2023"
}

@article{Moch:2004pa,
    author = "Moch, S. and Vermaseren, J. A. M. and Vogt, A.",
    title = "{The Three loop splitting functions in QCD: The Nonsinglet case}",
    eprint = "hep-ph/0403192",
    journal = "Nucl. Phys. B",
    volume = "688",
    pages = "101--134",
    year = "2004"
}

@article{Goyal:2025bzf,
    author = "Goyal, Saurav and Moch, Sven-Olaf and Pathak, Vaibhav and Rana, Narayan and Ravindran, V.",
    title = "{Soft and virtual corrections to semi-inclusive DIS up to four loops in QCD}",
    eprint = "2506.24078",
    archivePrefix = "arXiv",
    primaryClass = "hep-ph",
    reportNumber = "DESY-25-079",
    month = "6",
    year = "2025"
}

@article{Moch:2005ba,
    author = "Moch, S. and Vermaseren, J. A. M. and Vogt, A.",
    title = "{Higher-order corrections in threshold resummation}",
    eprint = "hep-ph/0506288",
    archivePrefix = "arXiv",
    reportNumber = "DESY-05-105, SFB-CPP-05-25, DCPT-05-60, IPPP-05-30, NIKHEF-05-010",
    doi = "10.1016/j.nuclphysb.2005.08.005",
    journal = "Nucl. Phys. B",
    volume = "726",
    pages = "317--335",
    year = "2005"
}

@article{DeRos:2026bcv,
    author = "De Ros, Lorenzo and Forte, Stefano and Ridolfi, Giovanni and Tagliabue, Davide Maria",
    title = "{Threshold resummation of rapidity distributions at fixed partonic rapidity}",
    eprint = "2601.04309",
    archivePrefix = "arXiv",
    primaryClass = "hep-ph",
    reportNumber = "TIF-UNIMI-2025-24",
    month = "1",
    year = "2026"
}

@article{de_ros,
    author = "De Ros, Lorenzo and Forte, Stefano and Ridolfi, Giovanni and Tagliabue, Davide Maria",
    title = "{Threshold resummation of rapidity distributions at fixed partonic rapidity}",
    eprint = "2601.04309",
    archivePrefix = "arXiv",
    primaryClass = "hep-ph",
    reportNumber = "TIF-UNIMI-2025-24",
    month = "1",
    year = "2026"
}

@article{AbdulKhalek:2022hcn,
    author = "Abdul Khalek, R. and others",
    title = "{Snowmass 2021 White Paper: Electron Ion Collider for High Energy Physics}",
    eprint = "2203.13199",
    archivePrefix = "arXiv",
    primaryClass = "hep-ph",
    reportNumber = "FERMILAB-PUB-22-125-QIS-SCD-T",
    month = "3",
    year = "2022"
}

@article{Hoschele:2013pvt,
    author = {H{\"o}schele, Maik and Hoff, Jens and Pak, Alexey and Steinhauser, Matthias and Ueda, Takahiro},
    title = "{MT: A Mathematica package to compute convolutions}",
    eprint = "1307.6925",
    archivePrefix = "arXiv",
    primaryClass = "hep-ph",
    reportNumber = "SFB-CPP-13-50, TTP13-27, LPN13-052",
    doi = "10.1016/j.cpc.2013.10.007",
    journal = "Comput. Phys. Commun.",
    volume = "185",
    pages = "528--539",
    year = "2014"
}

@phdthesis{Ablinger:2012ufz,
    author = "Ablinger, Jakob",
    title = "{Computer Algebra Algorithms for Special Functions in Particle Physics}",
    eprint = "1305.0687",
    archivePrefix = "arXiv",
    primaryClass = "math-ph",
    school = "Linz U.",
    month = "4",
    year = "2012"
}

@article{Vermaseren:1998uu,
    author = "Vermaseren, J. A. M.",
    title = "{Harmonic sums, Mellin transforms and integrals}",
    eprint = "hep-ph/9806280",
    archivePrefix = "arXiv",
    reportNumber = "FTUAM-98-7, NIKHEF-98-014",
    doi = "10.1142/S0217751X99001032",
    journal = "Int. J. Mod. Phys. A",
    volume = "14",
    pages = "2037--2076",
    year = "1999"
}

@article{Remiddi:1999ew,
    author = "Remiddi, E. and Vermaseren, J. A. M.",
    title = "{Harmonic polylogarithms}",
    eprint = "hep-ph/9905237",
    archivePrefix = "arXiv",
    reportNumber = "NIKHEF-99-005, TTP-99-08",
    doi = "10.1142/S0217751X00000367",
    journal = "Int. J. Mod. Phys. A",
    volume = "15",
    pages = "725--754",
    year = "2000"
}

@book{Ellis:1996mzs,
    author = "Ellis, R. Keith and Stirling, W. James and Webber, B. R.",
    title = "{QCD and collider physics}",
    doi = "10.1017/CBO9780511628788",
    isbn = "978-0-511-82328-2, 978-0-521-54589-1",
    publisher = "Cambridge University Press",
    volume = "8",
    month = "2",
    year = "2011"
}

\end{document}